\documentclass[10pt]{article}
\usepackage[utf8]{inputenc}
\usepackage{graphicx}
\usepackage[misc]{ifsym}
\usepackage{caption}
\usepackage{subcaption}
\usepackage{wrapfig, booktabs}
\usepackage{floatflt}
\usepackage{fancyhdr}
\usepackage[a4paper, total={6.3in, 9in}]{geometry}
\usepackage{fourier}
\usepackage{wrapfig}
\usepackage{float}
\usepackage{hyperref}
\usepackage{enumitem}
\usepackage{amsmath}
\usepackage{comment}
\usepackage[framemethod=TikZ]{mdframed}
\usepackage{tikz}
\usetikzlibrary{positioning}
\usepackage{xcolor}

\definecolor{Acol}{rgb}{0.11, 0.53, 0.93}
\definecolor{AMIPcol}{rgb}{0.11, 0.53, 0.93}
\definecolor{Bcol}{rgb}{0.6, 0.8, 0.2}
\definecolor{Ccol}{rgb}{1.0, 0.84, 0.0}
\definecolor{Dcol}{rgb}{1.0, 0.5, 0.0}
\definecolor{Aggcol}{rgb}{0.8, 0.15, 0.15}
\definecolor{gacol}{rgb}{0, 0.494, 0.725}
\mdfdefinestyle{MyFrame}{%
    linecolor=black,
    outerlinewidth=1pt,
    roundcorner=5pt,
    innertopmargin=\baselineskip,
    innerbottommargin=\baselineskip,
    innerrightmargin=20pt,
    innerleftmargin=20pt,}
\usepackage{multicol}
\usepackage[gen]{eurosym}
\usepackage[
backend=biber,
style=nature,
sorting=none
]{biblatex}

\hypersetup{
    colorlinks=true,
    linkcolor=blue,
    filecolor=magenta,      
    urlcolor=cyan,
}
\begin{document}
\begin{center}
{\Large\textbf{Through energy droughts: Hydropower’s ability to sustain a high output}}\\
\vspace{5mm} 
\end{center}
{\small \textit{Hanna Ek Fälth $^{1,3}$ \Letter, Fredrik Hedenus $^1$, Lina Reichenberg $^1$, Niclas Mattsson $^2$\\1 \hspace{0.4cm} Department of Space, Earth and Environment, Division of Physical Resource Theory, Chalmers University of Technology, Gothenburg, Sweden\\2 \hspace{0.4cm} Department of Space, Earth and Environment, Division of Energy Technology, Chalmers University of Technology, Gothenburg, Sweden\\\Letter \hspace{0.3cm} hanna.ek.falth@chalmers.se}}

\rule[0.8ex]{\linewidth}{1pt}


\subsection*{Summary}

Previous research has raised concerns about energy droughts in renewables-based energy systems. This study explores the ability of reservoir hydropower to sustain a high output and, thereby, mitigate such energy droughts. Using detailed modelling, we estimate that Swedish hydropower can sustain 67–92\% of its installed capacity for 3 weeks, with higher values possible in springtime. The variation of the sustained output, equivalent to the capacity of 3–4 Swedish nuclear reactors, under-scores the importance of understanding the potential output levels when devising strategies to counteract energy droughts. Moreover, we find that regulations imposed on the flows in river bottlenecks hinder higher sustained output levels. With the upcoming renewal of environmental permits for hydropower plants in Sweden, these findings provide valuable insights for policymakers. Furthermore, the sustained output capabilities demonstrated in this study challenge the prevalent simplified representations of hydropower in energy models, suggesting a need for more-sophisticated modelling approaches.

\subsection*{Keywords}
Hydropower, Modelling, Energy systems, Energy droughts, Dunkelflaute event, Resilience, Sustained output, Renewables



\newpage
\section{Introduction}

Variation management strategies are essential in renewables-based energy systems to ensure that the demand can be met at all times despite variations in wind and solar power production. Energy storage, expansion of transmission grids, demand-side management, and dispatchable generation technologies are the key strategies discussed in the literature \cite{sinsel_challenges_2020, kirkerud_power--heat_2017, rodriguez_transmission_2014, child_flexible_2019, lund_review_2015, johansson_impacts_2020, ruhnau_storage_2022, kondziella_techno-economic_2023, soder_review_2018}. Furthermore, reservoir hydropower has been argued to provide flexibility in renewables-based systems \cite{hirth_benefits_2016, dimanchev_role_2021, owolabi_robust_2022, liu_role_2019, zhao_importance_2023, phillips_metric_2020, thapa_factors_2022}. In this study, we analyse the ability of reservoir hydropower to provide capacity during a sustained period of low availability of supply relative to demand, a so-called \emph{energy drought}.

In contrast to many other variation management strategies, hydropower has been used in power systems for decades, and has long been recognised for its operational flexibility. In Sweden, hydropower has contributed significantly to system flexibility by providing diurnal production to follow load, seasonal storage capabilities, and grid stability. However, the shift toward renewable energy systems could alter the way in which hydropower is optimally utilised. A recent study by \textcite{oberg_evaluation_nodate} has revealed a notable shift in hydropower’s operational dynamics, in that the conventional daily production cycle is becoming less pronounced, influenced by the increasing integration of wind power into the electricity system. This trend suggests a re-evaluation of hydropower’s role, which may be transitioning from its traditional focus on meeting intra-day demand fluctuations to accommodating the variability of renewable energy sources on different time-scales.  

Energy systems that have a large share of renewables face numerous challenges related to variability, ranging from short-term issues such as maintaining frequency control and adapting to hourly load changes to persistent supply shortages due to factors such as low-wind periods or technology failures. Researchers have explored hydropower’s role in addressing these challenges across different time-scales. \textcite{phillips_metric_2020} have developed a framework to assess hydropower’s potential for enhancing short-term grid resilience after a disturbance, highlighting reservoir hydropower’s critical role in managing disruptions. \textcite{yang_burden_2018} have quantified the quality of short-term regulation of hydropower and the burden placed on generation equipment, and they have also evaluated burden relief strategies under different future variable renewable energy (VRE) scenarios. Extending the research to intra-day variations, \textcite{thapa_factors_2022} have examined the capacity of a cascaded reservoir hydropower system to meet daily demand peaks under various operational constraints, and they have identified some key factors influencing intra-day flexibility. However, hydropower’s potential to mitigate week-long energy droughts is poorly explored. Such droughts can result from, for instance, weather phenomena \cite{staffell_increasing_2018, li_brief_2021, raynaud_energy_2018, most_temporally_2024, otero_copula-based_2022, kapica_potential_2024}, technical failures in transmission lines \cite{stankovski_power_2023, fotis_risks_2023}, and emergency shutdowns of nuclear power plants \cite{ahmad_increase_2021, jeong_estimating_2022}. This study aims to fill this knowledge gap by evaluating hydropower’s ability to sustain a high output over extended periods, thereby assessing its capacity to counteract energy droughts.

Historical hydropower production data do not show instances of hydropower operating at near-maximum capacity for week-long periods. This could be due to either inherent operational limitations of the hydropower systems or the lack of economic incentives in the past. Therefore, rather than relying solely on historical data, employing a model becomes crucial to understanding hydropower’s ability to sustain high output levels over longer periods of time with strong incentives. However, the energy systems models used to date often simplify the representation of hydropower, neglecting key elements such as river network effects, water delay times, and head-dependent production, thereby over-estimating the flexibility of hydropower \cite{ek_falth_trade-offs_2023}. Both the International Energy Agency (IEA) and the National Renewable Energy Laboratory (NREL) advocate for more-detailed modelling that can accurately assess hydropower’s flexibility \cite{huertas-hernando_hydro_2017, stoll_hydropower_2017}. Overly simplified hydropower representations imply that hydropower can operate continuously at maximum output for as long as water is available in the reservoirs. However, is that the case in reality? This study aims to answer this question by employing a detailed model to assess hydropower’s sustained output capabilities and to offer insights into its potential to mitigate the impacts of energy droughts.

The importance of such an analysis extends globally, especially to regions that are heavily dependent upon reservoir hydropower, like South America, Canada, China, Central Africa, and parts of the USA and Europe. As emphasised above, a detailed representation of the hydropower system is essential for accurate evaluation of its flexibility. Thus, this paper focuses on Sweden as a case study, leveraging a detailed model of its hydropower infrastructure to examine hydropower’s ability to sustain a high level of output for 1–3 weeks. The installed hydropower capacity in Sweden is about 16 GW, and in 2022 about 70 TWh of the Swedish electricity generation originated from hydropower, corresponding to 41\% of the total electricity production \cite{noauthor_bruttoproduktion_nodate}. Our model covers 92\% of the hydropower capacity in Sweden, encompassing nine major rivers and incorporating around 240 reservoirs and power plants, each of which is equipped with 1–10 turbines. Further details of the model and coverage are provided in the \textit{Experimental procedure} section.  

The methodology that we use to represent hydropower in a detailed manner is derived from the approach outlined previously by \textcite{ek_falth_trade-offs_2023}. Briefly, it is an optimisation model that maximises profit for a set of hydropower plants subjected to deterministic spot prices over a full year. This model represents each component of the river systems, including reservoirs, flow paths with their respective flow times, and individual turbines with their respective performance curves. The head dependency of hydropower production is modelled, and environmental constraints, such as minimum flows and seasonally dependent limits on water levels, are also represented. Notably, the spillage representation is improved for this study to depict more accurately the bottlenecks in the rivers, which are shown to heavily influence hydropower’s ability to sustain output. 

To assess the ability of hydropower to sustain a high output during energy droughts, we simulate high-demand conditions by introducing constant, high market prices for periods of 1–3 weeks into the historical price profiles. This method allows us to mimic the economic signals that would trigger hydropower plants to operate at increased output levels, mimicking real-world energy droughts. By analysing the hydropower output during these high-price periods, we can enhance our understanding of hydropower’s potential role in addressing energy system challenges, such as prolonged energy shortages. Initially, we investigate the potential to sustain a high output and how it is affected by the upper limits imposed on flows in the river. Subsequently, we map the seasonal patterns to explore the relationship between sustained output variations and energy drought occurrences.


\section{Measuring sustained output} \label{sec:sustained_output}
To address energy droughts effectively, we require technologies that can meet the demand throughout these critical periods.
This study introduces two metrics to evaluate the ability to sustain a high output: \textit{Sustained capacity} and \textit{Sustained production}. 

\begin{itemize}
\item \textit{Sustained capacity} is defined as the consistent power output that hydropower plants can guarantee throughout a specified period. This metric is measured as a percentage of the maximum capacity of the plants.
\item \textit{Sustained production} is defined as the total electrical energy generated during a specific period. This metric is expressed as the percentage of the maximum possible production, i.e., the total production from hydropower plants running at maximum capacity throughout this period.  
\end{itemize}

\begin{figure}[H]
    \centering
    \includegraphics[width=0.49\linewidth]{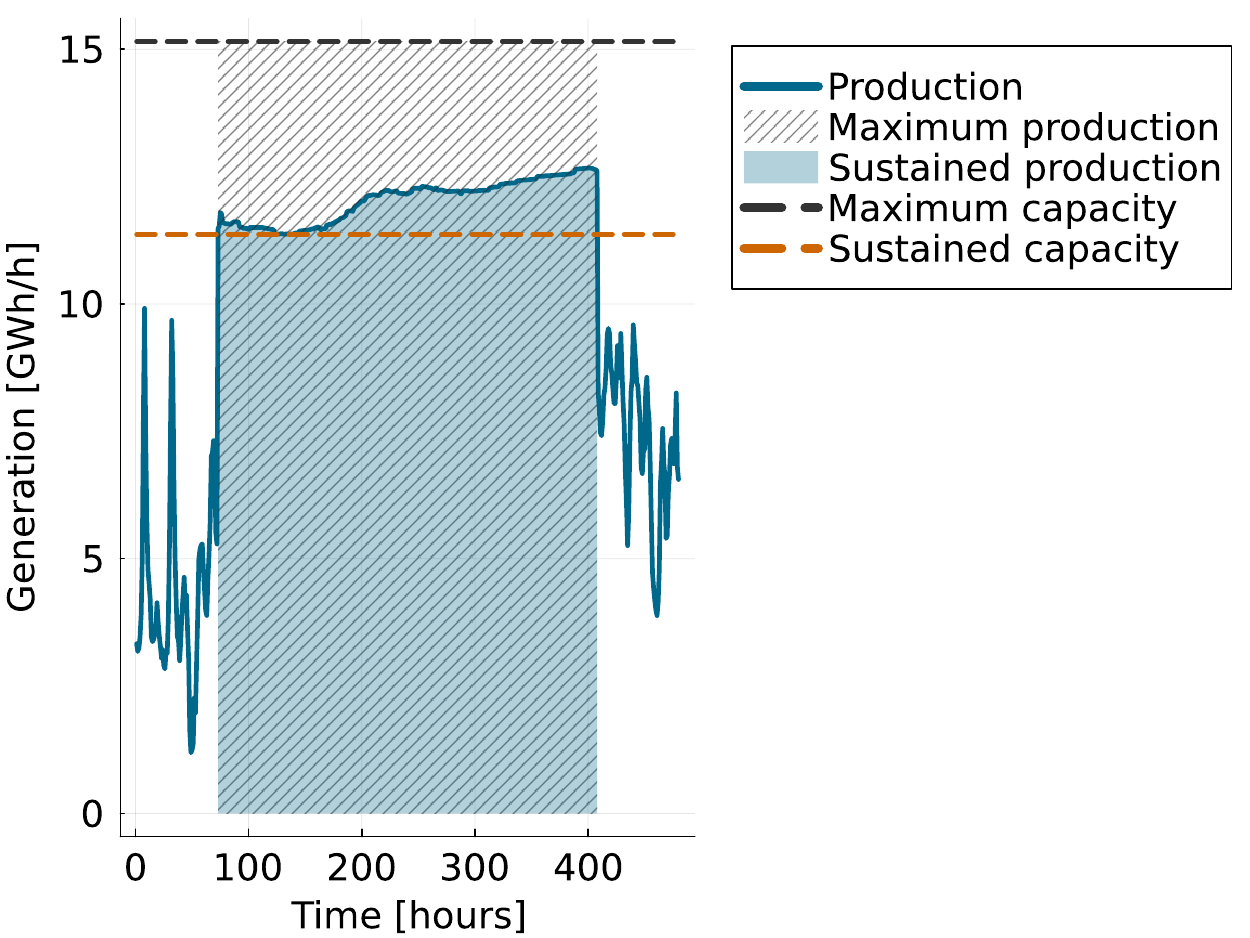}
    \caption{Illustration clarifying the definitions of the metrics of Sustained capacity and Sustained production. Shown are the levels of hydropower production over a 20-day period, and the metrics during a simulated 14-day energy drought.}
    \label{fig:defineSO}
\end{figure}

Figure \ref{fig:defineSO} shows the hydropower production profile over 20 days with a simulated 14-day energy drought in the middle, and shows the metrics of Sustained capacity and Sustained production applied to this example.  

\section{Results \& Discussion}

\subsection{Hydropower capacity can be sustained at high levels over several weeks}\label{sec:results:overall}
\begin{figure}[H]
    \centering
            \centering
            \includegraphics[width=0.49\textwidth]{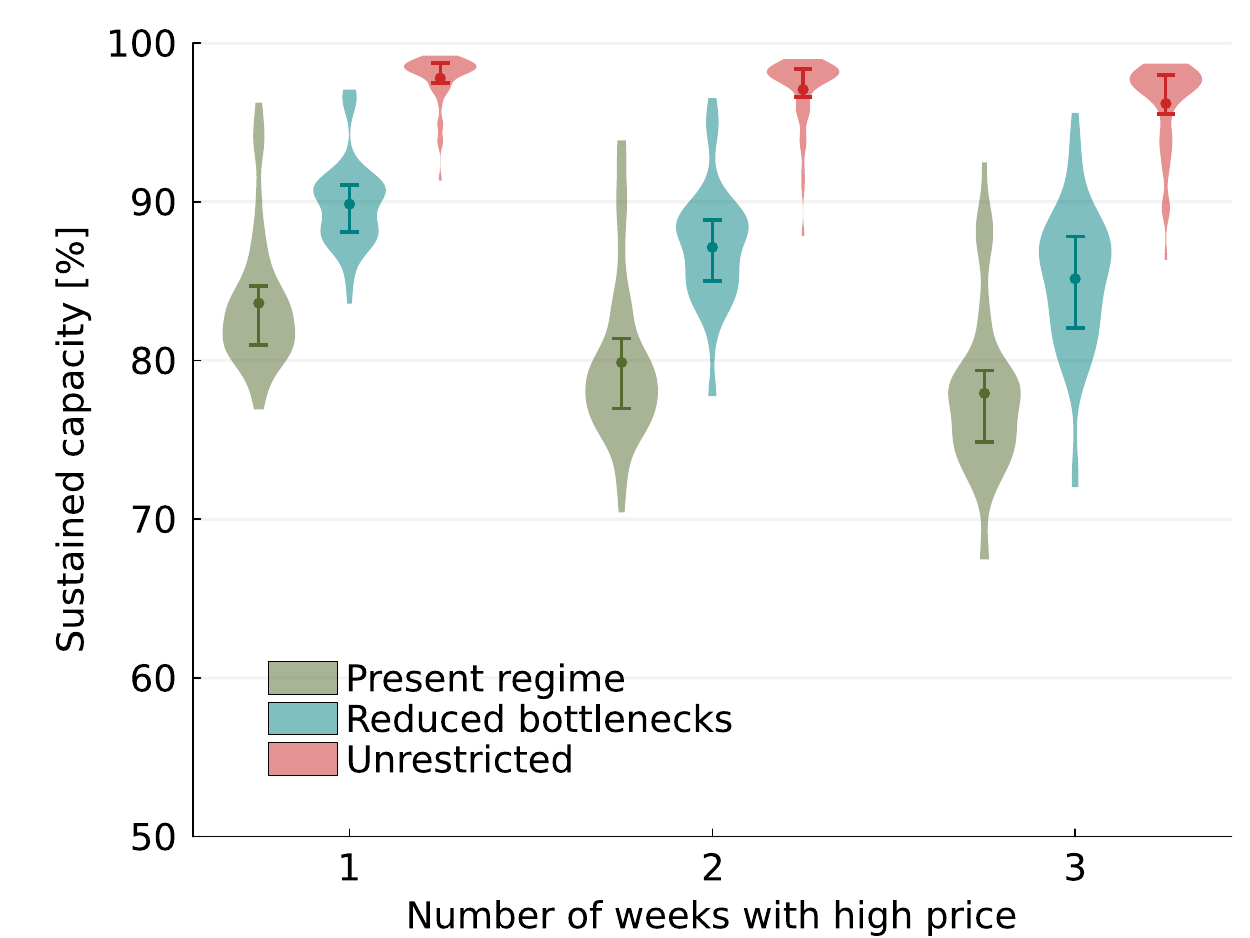}
    \caption{An illustration of the Sustained capacity of hydropower over a period of one to three consecutive weeks of high electricity prices for three different operational scenarios. Each violin shows the results for each month over four different years. The dots represent the mean values, while the lines extending from the dots indicate the 25th to 75th percentile range. The color-coding corresponds to the following operational scenarios: \textit{Present regime} with current regulations and infrastructure (green); \textit{Reduced bottlenecks} with a higher upper limit on spillage in bottlenecks (blue); and \textit{Unrestricted}, with unrestricted spillage (red). The different scenarios highlight the influences of operational constraints on the ability of hydropower to sustain a high output during energy droughts. The results shown are based on approximately 6,500 model runs. Refer to the \textit{Experimental procedure} section for details of the test set-up.}
    \label{fig:SC}
\end{figure}

Our findings reveal that hydropower can sustain high production levels over several weeks. Under current infrastructure and regulatory conditions, the Sustained capacity of Swedish hydropower ranges from 67\% to 96\%, depending on the time of year and the durations of high-price periods. On average, across all months spanning the period of 2016–2019, the Sustained capacity for a 1-week period was 84\%, decreasing to 78\% for a 3-week period. Regardless of the season or year, Swedish hydropower consistently maintained at least 67\% of its total capacity for up to 3 weeks. These results are illustrated in Figure \ref{fig:SC} for the \textit{Present regime} scenario, which showcases hydropower’s Sustained capacity over one to three consecutive weeks. The results obtained for Sustained production are very similar to those for Sustained capacity and are, therefore, not shown here, although they are provided in Figure \ref{fig:A:SO} in the \textit{\textit{Supplementary material}}. Furthermore, the results for the sustained output divided according to Nord Pool price areas in Sweden are presented in Figures \ref{fig:A:SC} and \ref{fig:A:SP} in the \textit{\textit{Supplementary material}}.

Further examination of Figure \ref{fig:SC} indicates that the variation in hydropower’s ability to sustain output over 1 week versus 3 weeks is less significant than the variation observed across different months and years, as highlighted by the magnitudes of the variability within each violin. This suggests that the timing of energy droughts plays a more critical role than drought duration in determining hydropower’s capacity to maintain output during these periods.

We have modelled and analysed 92\% of Sweden’s installed hydropower capacity, focusing on the largest and most flexible river systems. Since the remaining capacity primarily consists of plants with lower levels of operational flexibility, the results for sustained output should not be extrapolated to this remaining capacity. It is also worth noting that we assumed that all power plants would be available for production at all times. In reality, owing to planned and unplanned shutdowns for maintenance and failures, the actual sustained production may occasionally be lower than is indicated by our results.

Furthermore, the unpredictable natures of the inflow and prices pose challenges for production planning, limiting the flexibility of hydropower compared to the deterministic model’s idealised approach used in this study. For instance, deterministic models may completely empty or fill reservoirs with confidence because they know precisely when new water will arrive and when a high-price period will appear. This cannot be accomplished in practice, as it may lead to exceeding the upper and lower bounds of reservoir content due to incorrect inflow forecasts. We introduced the high-price period only 1 week after the initial modelled hour, to reflect more accurately the actual conditions and to reduce the over-estimation of flexibility from the deterministic model. This adjustment restricts the model’s ability to optimise the water levels before encountering the high-price period. Nevertheless, even with this modelling approach, using a deterministic model still somewhat over-estimates the possibilities for hydropower to sustain output.

\subsection{Adjusting the allowed water flows at bottlenecks enhances hydropower output during energy droughts}

The current regulatory regime for hydropower production in Sweden includes limitations on allowed water flow rates in different sections along the river. The regulations generally permit operators to run water through or bypass power plants up to the maximum flow capacity of the installed turbines, except during high-inflow periods when more-significant water flow rates are allowed or forced. Bottlenecks arise when the maximum flow capacity of the turbines in a power plant is lower than that of plants upstream or downstream, thereby impeding water transfer. For more details on bottlenecks, see the \textit{Experimental procedure} section. 

To explore the impacts of bottlenecks on sustained output, we examined two alternative scenarios. First, we analysed an \textit{Unrestricted} scenario, in which the upper limits on water flow rates were removed at each hydropower plant, while maintaining maximum turbine flows, thereby effectively increasing the spillage limits. Allowing unrestricted spillage throughout the river could raise the sustained output to >95\% on average for periods of up to 3 weeks (illustrated by the red violins in Figure \ref{fig:SC}), as compared to 77\% under current regulatory regimes (green violins in Figure \ref{fig:A:SC}). However, unrestricted spillage entails severe risks, such as flooding and erosion. Although unrealistic, this scenario indicates the impact of current water flow rate limits on sustained output and serves as an upper bound on the possibilities for sustained output.

Second, we evaluated a more targeted approach that allowed water flows to exceed the turbine flow capacities at identified bottleneck sites. In this scenario, which we refer to as the \textit{Reduced bottlenecks} scenario, we raised the flow limits in approximately 30\% of the power plants to match the maximum turbine flow rates of the upstream turbines, effectively increasing their water flow rate limits by an average of 20\%. The adjustments to the allowed water flows were made without increasing the maximum turbine flow capacities, thus only increasing the spillage limits. When we compared these revised flow rate limits to the maximum recorded inflow for each plant (including local and upstream contributions) during the period of 2016–2020, the limits increased from an average of 18\% to 22\% of the maximum inflow. Similarly, when compared to the mean inflow during the same period, the spillage limits increased from an average of 163\% to 194\% of the mean inflow. (See the \textit{Experimental procedure} section for more details on how and why flow limits were expanded at bottlenecks). In the \textit{Experimental procedure} section, adjustments to the flow rate limits, as shares of both the maximum and average inflows, are detailed in Figure \ref{fig:expansion_info}.

In this \textit{Reduced bottlenecks} scenario, Sustained capacity averages 90\% for 1 week and 85\% for 3 weeks (blue violins in Figure \ref{fig:SC}). This marked improvement over the \textit{Present regime} scenario demonstrates that adjusting spillage regulations at bottlenecks notably enhances hydropower’s performance during energy droughts. Nevertheless, while the benefits of increasing flow limits are clear, they must be carefully balanced against potential environmental and social impacts, which could have negative effects on the river ecosystems and the communities that depend on them.

Our analysis did not capture all the complexities related to spillage, since we were constrained by data limitations. Each power plant operates under individual permits that regulate releases of water. We generalised these regulations by setting the upper limits on water flows through the turbines or in spillways to be equal to the maximum installed turbine flow. This assumption is based on consultations with experts from power production companies and water regulation authorities. Furthermore, water flows are restricted during wintertime due to ice formation in rivers and the risk of mechanical issues, likely resulting in a reduced ability to sustain output during the colder months. Given the generalisation of permits and the omission of the ice-related problems, our results should be viewed as indicative of the potential for sustained high output in Swedish hydropower rather than definitive calculations.

Furthermore, it is important to note that achieving sustained high-level production in rivers fundamentally depends on maintaining high flow rates throughout the entire river system. Our approach focused on increasing the spillage limits to improve flow at bottlenecks. Another effective solution could be to expand the turbine flow capacity by installing additional turbines at these critical points. Such an expansion would enhance the river system’s ability to sustain high output and increase capacity while avoiding the increased annual energy losses associated with higher spillage.

\subsection{Higher sustained hydropower capacity can be obtained during periods of high natural inflow and high reservoir levels}\label{sec:results:months}

\begin{figure}[H]
    \centering
    \includegraphics[width=0.9\linewidth]{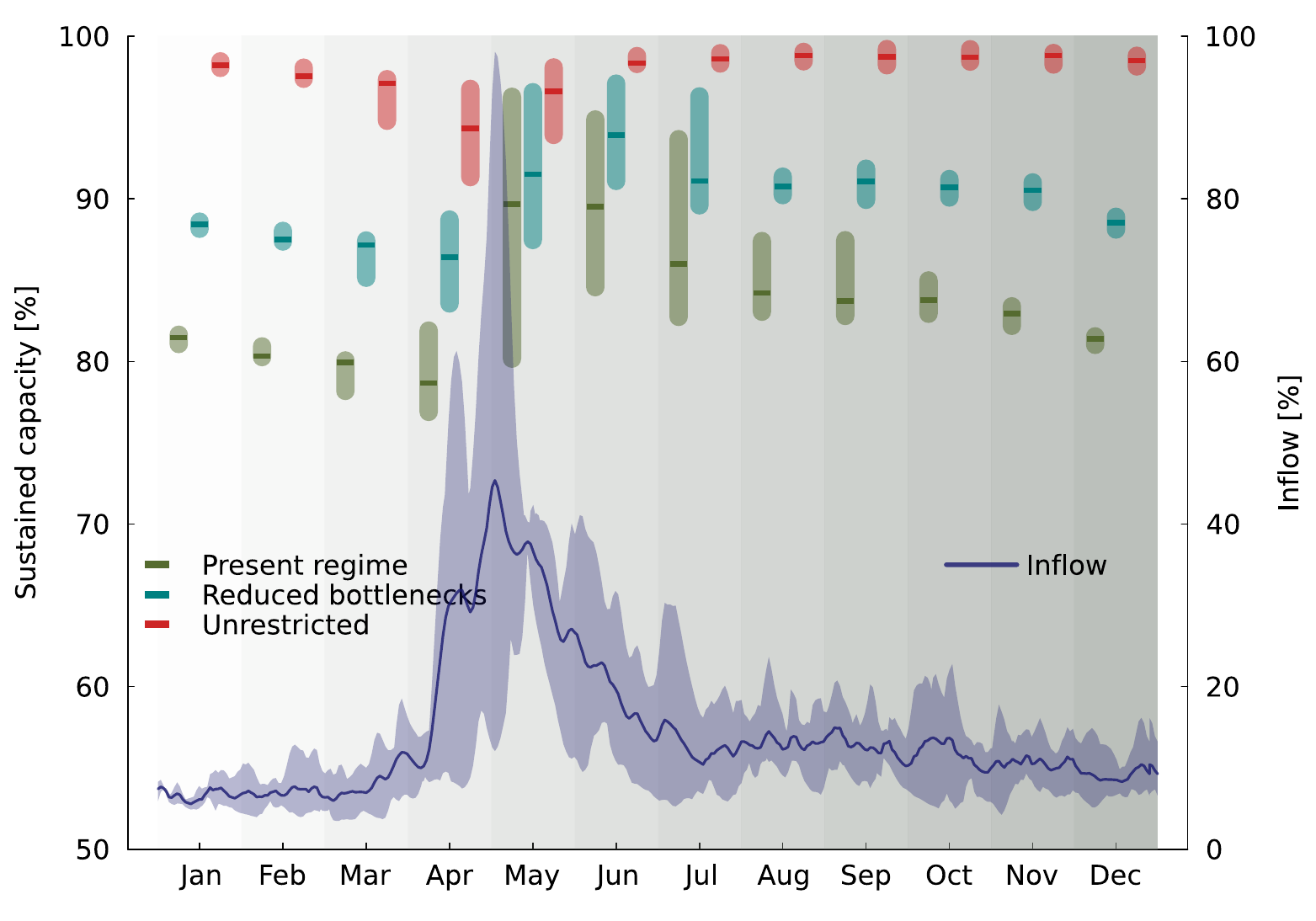}
    \caption{A comparison of the monthly Sustained capacity of Swedish hydropower under three operational scenarios and the natural inflow throughout the year, illustrating how today’s hydrological cycle influences hydropower’s ability to sustain a high output. The Sustained capacity values under the \textit{Present regime} (green), \textit{Reduced bottlenecks} (blue), and \textit{Unrestricted} (red) scenarios are shown in box plots, which display the range and mean values for each month. The line graph represents the total inflow as a percentage of the maximum recorded. The shaded area behind the line graph indicates the variability of the inflow for the modelled period of 2016–2020.}
    \label{fig:SC_months}
\end{figure}

Our findings reveal notable variability in the sustained output across the \textit{Present regime} and \textit{Reduced bottlenecks} scenarios, as illustrated by the green and blue violins, respectively, in Figure \ref{fig:SC}. Figure \ref{fig:SC_months} shows the 1-week Sustained capacity by month for three operational scenarios: \textit{Present regime}, \textit{Reduced bottlenecks}, and \textit{Unrestricted}. The \textit{Unrestricted} scenario consistently demonstrates sustained capacities >98\%, except from March to May. This decrease can be explained by current operational practices, whereby reservoir levels are typically at their lowest before the spring run-off due to intentional winter draw-downs to meet the high demand and to accommodate anticipated spring recharge. Consequently, diminished reservoir volumes limit the capacity for full output over 1 week, and reduced head heights in plants with significant reservoir level variability decrease the energy potential, thereby affecting the power output. This also explains the downward trend in Sustained capacity observed from November to March in both the \textit{Present regime} and \textit{Reduced bottlenecks} scenarios (Figure \ref{fig:SC_months}).

In addition, Figure \ref{fig:SC_months} charts the total natural inflow to the modelled rivers over the year as a percentage of the maximum observed, illustrating the correlation between inflow and Sustained capacity. A fluctuating inflow throughout the year affects the permitted spillage levels, as high-level spillage is only allowed during periods of naturally high inflow, such as during snowmelt or consistent rainfall, as detailed in the \textit{Experimental procedure} section. Consequently, the potential for high sustained output is greater during these high-inflow periods, as is evident in both the \textit{Present regime} and the \textit{Reduced bottlenecks} scenarios. This pattern arises from the design of the spillage constraints; natural inflow typically exceeds turbine capacity mainly in the springtime and occasionally in autumn, dictating spillage limits only during these periods. Allowing higher flows leads to a higher Sustained capacity because it enables water to bypass bottlenecks, and downstream plants can maintain high production levels using water from upstream reservoirs. For the same reasons, months with higher variability in terms of inflow also exhibit greater variability of Sustained capacity. Notably, the sustained output in April appears low despite there being a high inflow. This discrepancy is due to two factors: 1) the peak inflows typically occur at the end of April, whereas we introduced the high price period in the middle of the month; and 2) as discussed above, reservoir levels are generally lower before the spring run-off due to the high demand during wintertime and the anticipated recharge.

Thus, two primary factors determine the ability of hydropower to sustain a high hydropower output: 1) the water levels in the reservoirs; and 2) more critically, the permitted flow rates. These factors contribute to significant variability in hydropower’s ability to sustain a high output throughout the year and across different years, as illustrated in Figure \ref{fig:SC_months}. However, it is important to consider that future changes in inflow patterns driven by climate change, as well as adjustments made to reservoir management in response to these inflows and evolving energy demands, will alter the yearly differences in hydropower’s potential to sustain output.

\subsection{High Sustained capacity creates annual energy losses}
\begin{figure}[H]
    \centering
    \includegraphics[width=0.49\textwidth]{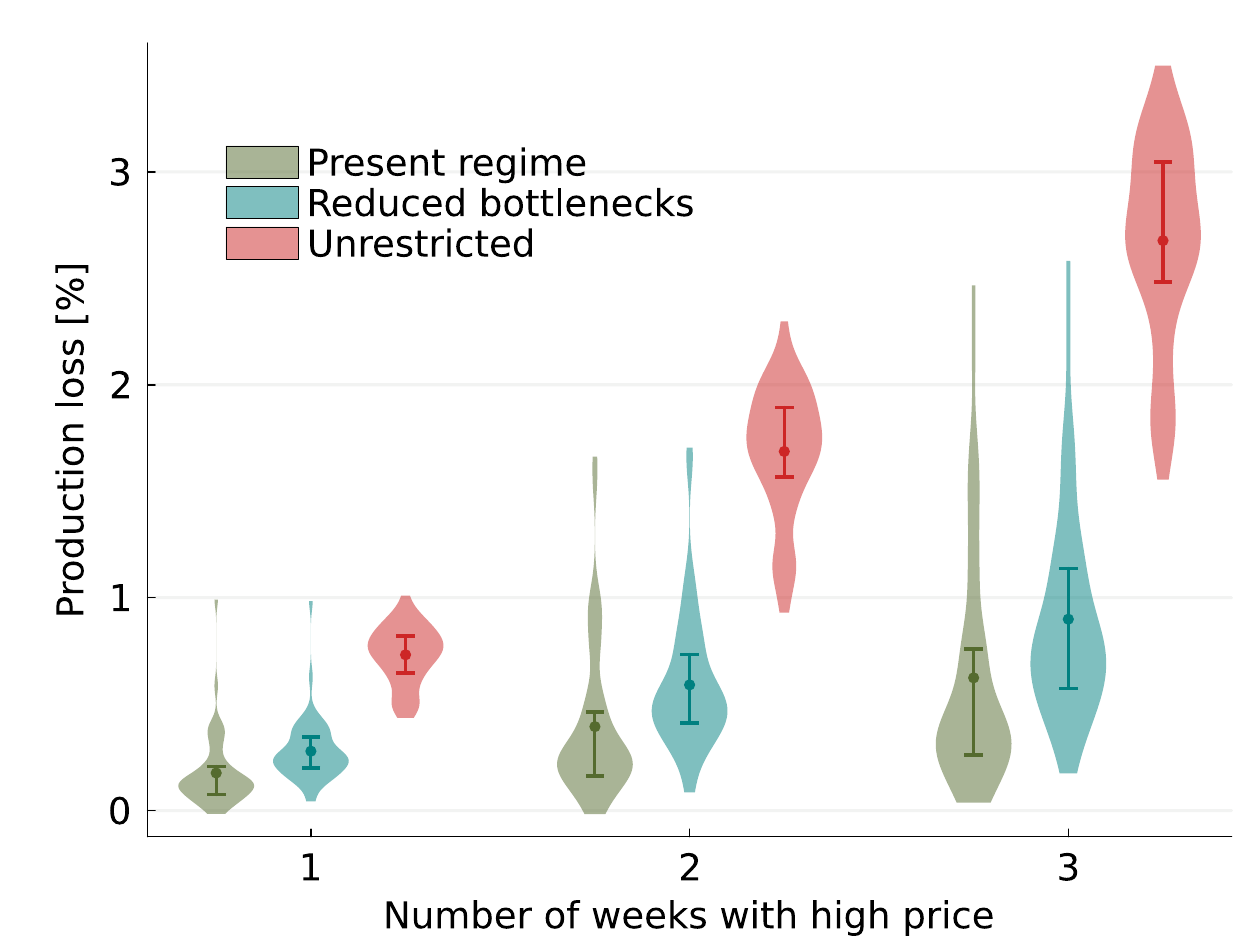}
    \caption{Losses in yearly production associated with sustaining high output levels, comparing the yearly production obtained with elevated market prices with the yearly production obtained with historical price levels. Each violin represents the results for each month over 5 different years. The dots represent the mean values, while the lines extending from the dots indicate the 25th to 75th percentile range. The color-coding corresponds to the following operational scenarios: \textit{Present regime}, with the current regulations and infrastructure (green); \textit{Reduced bottlenecks}, with a higher upper limit on spillage in bottlenecks (blue); and \textit{Unrestricted}, with unrestricted spillage (red).}
    \label{fig:Loss}
\end{figure}

Achieving a high sustained hydropower capacity inevitably reduces the annual energy output, primarily due to increased spillage at bottlenecks and, to a lesser extent, reduced turbine efficiencies at maximum flows. Figure \ref{fig:Loss} illustrates the annual production losses that are entailed by a high Sustained capacity, contrasting the outputs from scenarios involving elevated market prices with scenarios that have historical price levels. In the \textit{Unrestricted} scenario (Figure \ref{fig:SC}), sustaining an average capacity of 96\%–98\% results in a loss of about 0.8\% of the annual production for each week of sustained high output, as depicted by the red violins in Figure \ref{fig:Loss}. Meanwhile, with the current regulations and infrastructure, an average Sustained capacity of 78\%–84\% is achievable (as shown in the \textit{Present regime} scenario in Figure \ref{fig:SC}), with a corresponding lower loss of about 0.2\% of yearly production per week.

In summary, sustaining a high output from hydropower, so as to counteract energy droughts, leads to some energy losses, amounting to 0.2\%–0.8\% of annual production for each week of sustained output. This should be considered within the broader context of energy system needs. While losing renewable energy might appear to be negative in terms of the pursuit of a transition to a carbon-neutral energy system, the flexibility provided by hydropower is critical for renewables-based energy systems. It is generally more challenging to find renewable sources that offer this level of flexibility than to generate renewable energy in bulk.

\subsection{What does hydropower’s Sustained capacity, as measured in this study, imply for the management of energy droughts?}
Our analysis demonstrates that Swedish hydropower can deliver between 67\% and 96\% of its installed capacity during critical periods of 1–3 weeks. This capability under-scores the significant potential of hydropower to ensure the energy supply during multi-week energy droughts. However, evaluating the adequacy of this sustained output requires the consideration of several key factors.

First, the relevance of sustained hydropower output depends on its share of the energy mix. In Sweden, the lowest observed sustained output of 67\%, approximately 10 GW, can fulfil about 40\% of the peak electricity demand and 68\% of the average electricity demand. At the higher end, a Sustained capacity of 96\%, equivalent to 14.5 GW, could meet almost 60\% of the peak demand and almost 100\% of the average electricity demand. These values highlight the significant role that hydropower plays in enhancing the resilience of a renewable energy system, although they also point to the need for additional energy sources during severe energy droughts. The broad range of observed sustained outputs, with a 4.5 GW difference between the highest and lowest outputs – equivalent to the capacity of three to four Swedish nuclear reactors – underlines the need to evaluate carefully hydropower’s sustained output capabilities when developing strategies to counteract energy droughts.

Moreover, Sweden’s energy system's interconnection with the rest of Europe adds complexity to the assessment. While this interconnectivity often acts as a buffer against local shortages by enabling trade in energy, it also poses risks when energy droughts occur simultaneously across the continent, thereby amplifying the demand for reliable back-up solutions. 

Second, the timing and seasonality of the hydropower output relative to energy droughts are crucial. Energy droughts are more likely to occur during the wintertime in today’s European electricity system \cite{most_temporally_2024, li_brief_2021, otero_copula-based_2022}, and this trend may intensify with climate change \cite{kapica_potential_2024}. Moreover, historical data indicate that the winter months (December to February) are the periods of peak demand in Sweden \cite{noauthor_topplasttimmen_2024}, due to the use of electric heating. Our findings reveal that the ability of Swedish hydropower to sustain a high output is weakest during the winter months (November to March). This observation aligns with periods identified in recent literature as most susceptible to energy droughts. Unfortunately, this means that the seasons with the highest risk of energy shortages coincide with the periods during which there is a reduced capacity for hydropower to deliver high outputs.

\subsection{Understanding hydropower’s role in future energy systems}

While our study does not directly evaluate the role of hydropower in future energy systems by integrating a detailed representation within an energy system model, it offers valuable insights that can facilitate such assessments in future research. Our findings can be used to establish more realistic assumptions regarding hydropower’s ability to sustain a high output, compared to those used in the simplified representations that are commonly found in existing energy system models.

Energy system model studies have shown that reservoir hydropower can provide important flexibility in renewables-based systems \cite{hirth_benefits_2016, dimanchev_role_2021}. However, these models often assume that hydropower’s Sustained capacity is 100\% as long as there is water in the reservoirs. Our research indicates that the actual capacity to sustain high output can sometimes be as low as 67\%. Models that over-simplify hydropower’s capabilities may thus overestimate the resilience provided by hydropower and underestimate the need for additional assets that can provide backup during energy droughts. 

In contrast, some models that explore renewable energy systems use historical hydropower data to parameterise their representations. Such data have not historically shown the high sustained outputs that our study suggests are possible, due to the lack of economic incentives for such performance in previous energy systems. Therefore, relying on historical data to predict hydropower’s operation in future energy systems will likely under-estimate its flexibility.

The influences of economic incentives on sustained output are significant, as sustaining a high output involves a trade-off between maximising output during high-price periods and saving water for hours with lower prices. In addition, as we have shown, sustaining a high output entails a reduction of annual production, adding to this trade-off. We conducted a sensitivity analysis to determine whether the price level used in our study (5,000 SEK/MWh, approximately 430 €/MWh) effectively encourages maximising the sustained output during high-price periods. This analysis demonstrated that even at a substantially higher price of 50,000 SEK/MWh, the sustained output did not increase, thereby affirming that our chosen price point efficiently motivates Swedish hydropower plants to reach maximum output during these critical periods, given that prices during the remainder of the year align with the historical levels recorded from 2016 to 2020. Details of this sensitivity analysis are depicted in Figure \ref{fig:A:Sensitivity} in the \textit{Supplementary material}.

Flow rate limitations in the rivers are crucial when considering periods of elevated market prices. Interestingly, limits on flow rates had a marginal impact when only historical prices were considered, as in the past there were limited incentives for hydropower plants to sacrifice water to sustain high output. This represents a critical take-away message with respect to the use of historical data for model validation: factors that may seem to be irrelevant under the conditions of current energy systems and price structures could take on importance as new incentives emerge in future energy systems.

\subsection{Policy implications and future research}
Sweden is initiating a comprehensive process of reissuing environmental permits for approximately 2000 hydropower plants across the country, starting in Year 2024 \cite{noauthor_moderna_nodate, regeringskansliet_omprovning_2022}. Our study highlights the critical role of bottlenecks in influencing hydropower’s ability to sustain high output and, consequently, the hydropower sector’s capacity to mitigate energy shortages during periods of low wind or other critical phases for the energy system. While discussions surrounding the new environmental permits often revolve around minimum flows, the present study underlines the importance of considering upper flow limits. We highlight the potential to enhance hydropower’s ability to sustain high output through updates to environmental permits, particularly by allowing increased spillage or maximum turbine flows. However, these considerations must be carefully balanced against the risks for significant environmental consequences.

Given that hydropower constitutes a significant share of the energy system in many parts of the world, it is important to understand whether hydropower can sustain output at, for example, 50\% or close to 100\%, so as to evaluate the need for other types of back-up assets to ensure the supply during energy droughts. However, generalising the results of this study to other hydro-rich nations requires careful consideration, as each region presents unique technical and environmental challenges that affect the ability to sustain high output levels. Therefore, further studies of other regions are warranted to generate more robust conclusions about the broader applicability of our findings. 



\section{Experimental procedure}\label{sec:method}

\subsection{Resource availability}

\subsubsection{Lead contact}
For inquiries regarding the paper’s content, data, and model, please contact the lead author: Hanna Ek Fälth (hanna.ek.falth@chalmers.se).

\subsubsection{Materials availability}
No materials were used in this study.

\subsubsection{Data and code availability}
The model, written in Julia, utilised for the analysis in this study is available on GitHub \cite{noauthor_hannaekfalthrivermodelsweden_nodate}. Some of the data used as model input for this study could not be shared due to the confidentiality of the data. However, alternative data that can be used to run the model are provided. Licensing information can be found within the repository.

\subsection{Model overview}\label{sec:method:model}
The goal of this work was to determine the maximum possible capacity that hydropower plants within a river and across multiple rivers can sustain simultaneously for 1–3 weeks in the presence of strong economic incentives. To this end, we conducted stress tests using a detailed hydropower optimisation model that maximises profits \cite{noauthor_hannaekfalthrivermodelsweden_nodate}. These stress tests involved modifying the historical price profiles to include constant, high market prices over one, two or three consecutive weeks, simulating the economic signals that would prompt hydropower plants to operate at high output levels, akin to real-world scenarios of energy droughts.

Our modelling approach is based on the detailed hydropower model for rivers with multiple hydropower plants that has previously been comprehensively described as model B:L \cite{ek_falth_trade-offs_2023}. This linear optimisation model maximises profit for a set of hydropower plants throughout the year, with an hourly resolution, given deterministic inflows and spot prices for each river system. The model represents all the reservoirs, waterways, flow times, and individual turbines at each power plant. Turbine efficiencies are modelled using piece-wise linear approximations of actual non-linear efficiency curves, while the head-dependent production is linearised via a Taylor approximation. Furthermore, the model incorporates environmental constraints, such as minimum flow requirements and seasonally dependent minimum and maximum water levels. To ensure the accuracy of the linearised model for evaluations of sustained output, we conducted a sensitivity analysis using a full non-linear model (described as model A in \cite{ek_falth_trade-offs_2023}). This sensitivity analysis confirmed that the linearised model used in the present study accurately replicates the results regarding the ability to sustain high output levels, as compared to using a full non-linear model.

In addition to the technical features in the original model described in \cite{ek_falth_trade-offs_2023}, we implemented more realistic constraints on the allowed water flow at each hydropower plant. These constraints are dictated by technical capacities and ecological and social considerations, thus enhancing the model’s realism. Moreover, as demonstrated in this study, the constraints on water flow rates significantly impact the potential for sustained hydropower output, underscoring their importance in accurately addressing the main research question. Further details of these water flow limitations are presented in the section on \textit{Water flow limitations and scenarios}.

We have modelled hydropower in nine major rivers in Sweden\footnote{Luleälven, Skellefteälven, Umeälven, Indalsälven, Ångermanälven, Ljungan, Ljusnan, Dalälven and Göta älv (Klarälven + the stretch below Vänern)}, accounting for about 15 GW or 92\% of the nation’s hydropower capacity, as reported in the Year 2022 official statistics \cite{noauthor_bruttoproduktion_nodate}. Table \ref{tab:coverage} presents the geographic coverage of the modelled hydropower, segmented by Nord Pool price area.

\begin{table}[H]
    \centering
    \begin{tabular}{c|c|c|c|c|c}
                                                                & Total & SE1  & SE2  & SE3  & SE4  \\ \hline
Share [\%]                                                      & 92    & 98   & 99   & 72   & 0    \\ \hline
Modelled capacity [GW]                                          & 15.13 & 5.24 & 7.99 & 1.90 & 0    \\ \hline
Installed capacity [GW] \cite{noauthor_bruttoproduktion_nodate} & 16.40 & 5.36 & 8.08 & 2.64 & 0.32 \\ \hline
    \end{tabular}
    \caption{Modelled shares of Swedish hydropower capacity, detailing the percentages of total installed capacity in Sweden, both overall and according to Nord Pool price area.}
    \label{tab:coverage}
\end{table}

\subsection{Data overview}
For the parameterisation of our model across the nine rivers, we employed an extensive data-set spanning from Year 2016 to Year 2020. This data-set includes the hourly production levels, water levels at both the intake and outlet, turbine discharge rates, and spillage across the 240 reservoirs and power plants included in our analysis. All the companies that own these facilities provided the data under confidentiality agreements.

The input data for the parameterised model include the spot prices for each bidding area from Year 2016 to Year 2020, sourced from Nord Pool, and the hourly site-specific water inflow data for all rivers, as provided by the Swedish Meteorological and Hydrological Institute (SMHI) and the Water Regulation companies (Vattenregleringsföretagen). In addition, the hourly intake water levels were used to determine the start and end reservoir levels for each model run.

\subsection{Test design}\label{sec:method:tests}

\subsubsection{Representing energy droughts}
To evaluate the ability of hydropower to sustain high outputs during periods of energy droughts, we simulated conditions of high demand by incorporating constant high market prices into historical price profiles. This technique enables us to emulate the economic signals that would prompt hydropower plants to operate at increased output levels, mirroring real-world scenarios of energy droughts.

\begin{figure}[H]
    \centering
    \includegraphics[width=0.5\linewidth]{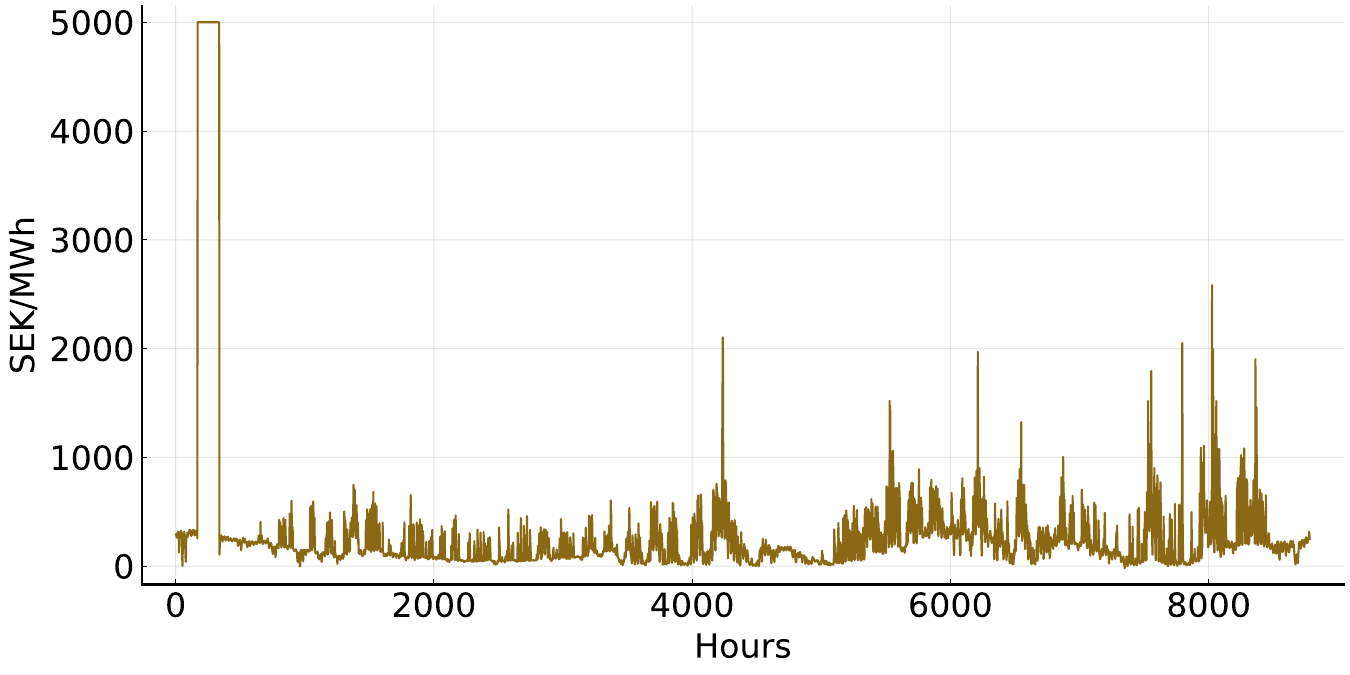}
    \caption{Example of a price profile with an introduced high-price period of 1 week in region SE3 in January 2020.}
    \label{fig:priceexample}
\end{figure}

We supplied the detailed optimisation model, previously described, with modified historical price profiles from 2016 to 2020. These profiles were adjusted to include one, two or three consecutive weeks with a constant high price of 5,000 SEK/MWh (approximately 430€/MWh). This pricing level was selected as a substantial incentive for high production levels; for reference, the peak price on the Swedish spot market in Year 2023 was 3,760 SEK/MWh. Figure \ref{fig:priceexample} demonstrates an example of a 1-week, high-price period introduced in January 2020 for the SE3 area.

We conducted a sensitivity analysis to confirm that the price level of 5,000 SEK/MWh effectively motivates maximum sustained output during these high-price intervals. The findings, detailed in Figure \ref{fig:A:Sensitivity} in the \textit{Supplementary material}, reveal that even a significantly higher price of 50,000 SEK/MWh did not increase the sustained output. This result confirms that our chosen price level effectively encourages Swedish hydropower plants to maximise their outputs during critical high-demand periods, assuming that prices during the remainder of the year align with the historical levels recorded between 2016 and 2020.

We utilised a model that has perfect foresight, incorporating deterministic water inflows and spot prices. To reduce the risk of over-estimating the sustained outputs – a consequence of using a model that is capable of precisely planning water releases and levels with complete foresight – we strategically introduced the high-price period just 1 week after the initial modelled hour. This approach restricts the model’s ability to optimise water levels in advance, thus enhancing the realism of our results for sustained output.

\subsubsection{Model runs}
To investigate how timing influences the sustained output capabilities of Swedish hydropower, we carried out an extensive series of model runs. Across each of the nine rivers, we executed 36 full-year model runs for each year between 2016 and 2019. These 36 runs (3 x 12) consisted of: one run with one high-price week; one run with two high-price weeks; and one run with three high-price weeks, repeated for each month. Thus, we modelled the period of January 2016 to January 2017 with a high-price period in January of one, two, and three consecutive weeks separately, and replicated this for all months and years up to December 2019 to December 2020. This methodological approach enabled us to assess how the timing of energy drought periods affects sustained production levels, as influenced by variations in inflow and initial reservoir levels.

To evaluate the potential losses in annual hydropower production due to a sustained high output, we conducted additional model runs using historical spot prices as baseline cases. These base cases consisted of 12 full-year model runs for each year between 2016 and 2019 (one for each month) to match the high-price week runs. These model runs allowed us to compare the annual production levels from the base case runs with those from the high-price periods, thereby quantifying the loss in annual production resulting from a sustained high output.

Maximum turbine capacities are not a direct input parameter in our model; they are inferred from the maximum turbine flows, varying head levels derived from the intake and outlet levels, and the configuration of the turbine efficiency curves. To determine the maximum capacity for each river, we ran the model with a high-price period that lasted just 1 hour and permitted any optimal initial reservoir levels. Establishing the maximum capacity enabled a meaningful comparison between the Sustained capacity during high-price weeks and the maximum capacity.

Furthermore, to examine how regulations imposed on water flow limitations impact the ability to sustain a high output, we analysed three scenarios concerning allowed water flows, as further explained in the \textit{Water flow limitations and scenarios} section.

In total, we conducted close to 6,500 model runs, considering five different price profiles (1-, 2-, and 3-week high-price runs, base case runs, and maximum capacity runs) for each year, for 4 years, across 12 months and nine rivers for three different spillage scenarios.

\subsection{Water flow limitations and scenarios}
In Sweden, each hydropower plant operates under a permit that specifies the allowed water flows. These water flow limitations can create bottlenecks at specific points along the river, significantly affecting the abilities of the hydropower plants in that river to sustain high outputs, as demonstrated by the findings of this study. This section provides an overview of how we define river bottlenecks and how they impact the ability to sustain high output levels. It further explains how water flow limitations were integrated into our model and details the three scenarios analysed in this study.

\subsubsection{Bottlenecks affecting the ability of hydropower to sustain a high output} \label{sec:bottlenecks} 
River flow rates typically increase as one moves downstream, due to the convergence of tributaries as the river progresses towards the sea. However, the larger reservoirs in Swedish river systems, which serve as substantial seasonal water storage facilities, are predominantly located upstream. In contrast, some downstream reservoirs can only hold enough water for a few days of full-capacity generation. Therefor, sustaining high production levels at downstream plants for prolonged periods requires the release of water from these larger upstream reservoirs, so as to compensate for the limited capacities of smaller downstream reservoirs.

\begin{figure}[H]
    \centering
    \includegraphics[width=0.15\textwidth]{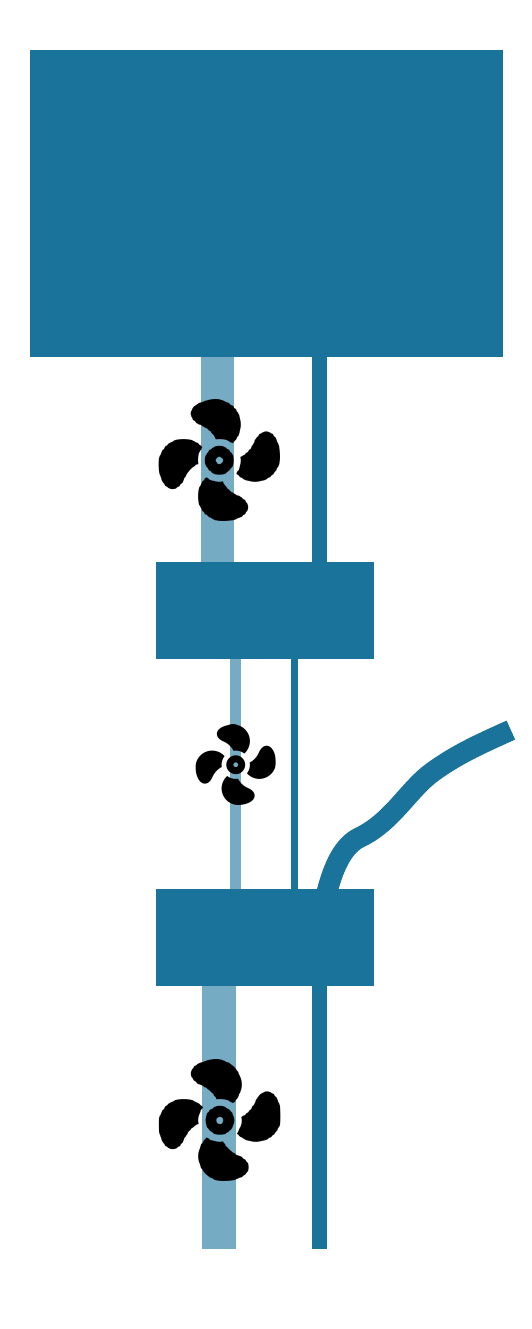}
    \caption{Conceptual representation of a river system, illustrating the concept of bottlenecks that can exist at various points along a river. The diagram depicts a series of reservoirs and turbines, with a large upstream reservoir followed by smaller ones downstream. It highlights how varying turbine capacities can create bottlenecks, particularly when a lower-capacity turbine is situated between higher-capacity turbines. Given the present regulatory frameworks and infrastructure, whereby the allowed flow through a plant (either through the turbine or in the spillway) generally is limited to the maximum flow capacity of the plant’s turbines, these bottlenecks diminish the ability to sustain high levels of hydropower production.}
    \label{fig:bottleneck}
\end{figure}

Bottlenecks arise where the allowed water flow rate of one plant is lower than others upstream or downstream, impeding water transfer. Figure \ref{fig:bottleneck} conceptually illustrates such a bottleneck. Consider a river system with a large upstream reservoir connected to a turbine and spillway with high-flow capability. Further downstream lies a smaller reservoir with a turbine and spillway that is capable of lower flows, followed by another small reservoir that is equipped with a high-flow turbine because it receives additional water from a tributary. The most-downstream turbine cannot sustain maximum capacity without rapidly depleting its reservoir, because even if ample water remains upstream, it cannot bypass the bottleneck (the middle turbine) quickly enough. Moreover, the most-upstream turbine cannot run at full capacity indefinitely without risking an overflow at the intermediary reservoir.

\subsubsection{Flow limits in the model}
The permit under which hydropower plants operate can limit the amount of water that they can release through turbines or in spillways, as well as the rate at which they can change water flows. For instance, some power plants can operate with so-called \textit{short-term regulation}, i.e., frequently changing the water flows withdrawn from their closest reservoir, while others do not.

Since some plants have the right to conduct short-term regulation, whereas some can only change their water flows at a limited rate of change, we introduced two variables to represent the flow constraints in our model: long-term spillage, $L_{t,p,p2}$; and short-term spillage, $S_{t,p,p2}$. The long-term spillage has the upper bound $l_{t,p,p2}$, as in Equation (\ref{eq:long}). The right to regulate water in the short term includes both spillage and the water released in the turbines. Thus, we set the sum of the short-term spillage, $S_{t,p,p2}$, and the turbine discharge, $D_{t,p,p2}$, to be below the upper bound for short-term regulation, $s_{t,p,p2}$, as in Equation (\ref{eq:short}). Index $t$ represents time, and the indices $p$ and $p2$ represent the passage from plant $p$ to downstream plant $p2$.    

\begin{equation}\label{eq:long}
    L_{t,p,p2} <= l_{t,p,p2} 
\end{equation}
\begin{equation}\label{eq:short}
    S_{t,p,p2} + D_{t,p,p2} <= s_{t,p,p2} 
\end{equation}

To restrict the rate of change in long-term spillage, we introduced constraints that limit the increase and decrease of long-term spillage from hour to hour. The parameter $i$ represents the allowed increase each hour, and $d$ represents the allowed decrease each hour.  

\begin{equation}
    L_{t,p,p2} <= L_{t-1,p,p2} + i_{t,p,p2}  
\end{equation}
\begin{equation}
    L_{t,p,p2} >= L_{t-1,p,p2} - d_{t,p,p2}  
\end{equation}

In our scenarios, we applied either short-term spillage or long-term spillage to each plant. In reality, it could be that one is allowed to release water above the limit for short-term regulation if that water release is carried out with slow changes to the flow. However, allowing for both long-term and short-term spillage in our model requires some model development since the current implementation would lead to the possibility of having short-term spillage on top of the long-term spillage, thereby changing the flow rates faster than is allowed above the upper limit for short-term regulation.   

\subsubsection{Scenarios}
To explore the impacts of bottlenecks on sustained output, we examined the following three scenarios with different upper bounds on spillage: \textit{Present regime}, \textit{Reduced bottlenecks}, and \textit{Unrestricted}. The scenarios are briefly explained below, and the entire code for all scenarios is available in the model at \cite{noauthor_hannaekfalthrivermodelsweden_nodate}.\\

\noindent \textbf{Present regime scenario} \\
Sweden’s current regulatory framework for hydropower production involves the issuing of specific permits for each plant. These permits typically impose restrictions on permissible water flow rates at specific points along the river. Some power plants can operate with short-term regulation, while others do not. For instance, reservoirs that are not directly linked to power plants frequently lack such permissions. Based on discussions with specialists from power production companies and water regulation authorities in Sweden, we conclude that power plants that are authorised to engage in short-term regulation are generally permitted to adjust flows rapidly between any minimum flow requirement and the maximum capacity of their installed turbines. In this study, for the ’Present regime scenario’, we have assumed that all reservoirs that are connected directly to power plants are entitled to regulate their water flows on a short-term basis. We have further assumed that flows that exceed the maximum capacity of the installed turbines, plus any mandatory minimum spillage requirements, are not permitted except during periods when the inflow exceeds the turbine’s maximum flow capabilities. Consequently, for these reservoirs, the upper limit on short-term regulation [$s$ in Equation (\ref{eq:short})] is set each hour by the greater of the maximum flow capacity of installed turbines or the present inflow to that reservoir. Furthermore, the upper limit on long-term spillage [$l$ in Equation (\ref{eq:long})] is set at zero.

For reservoirs that are not directly connected to a power plant, we set the short-term regulation [$s$ in Equation (\ref{eq:short})] to zero and permit long-term spillage [$l$ in Equation (\ref{eq:short})] every hour, equal to either the maximum inflow recorded that calendar month during the period of 2016–2020 or the mean inflow for the same period, whichever is greater during that hour.\\ 

\begin{figure}[H]
    \centering
        \begin{subfigure}[b]{0.49\textwidth}
            \centering
            \includegraphics[width=\textwidth]{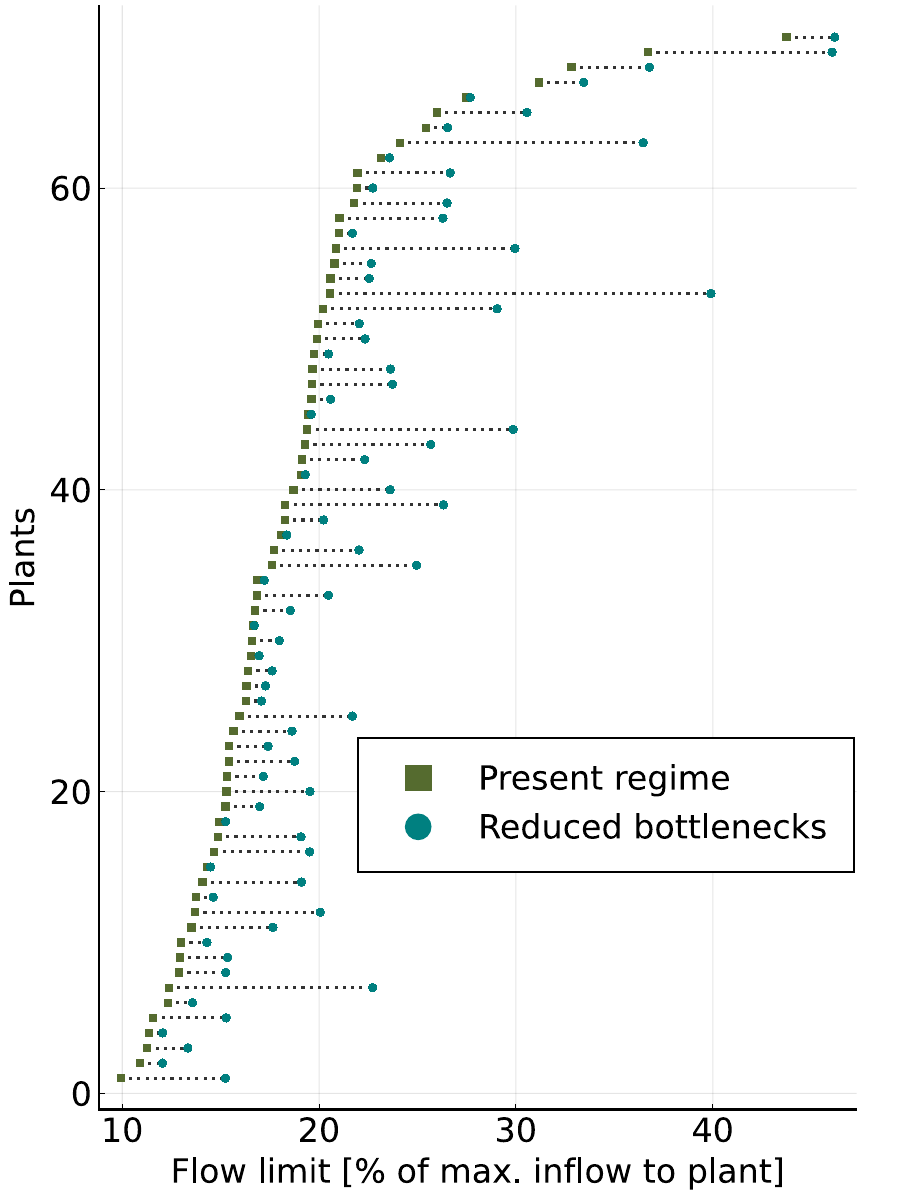}
            \label{fig:expansion_info_max}
        \end{subfigure}
        \hfill
        \begin{subfigure}[b]{0.49\textwidth}
            \centering
            \includegraphics[width=\textwidth]{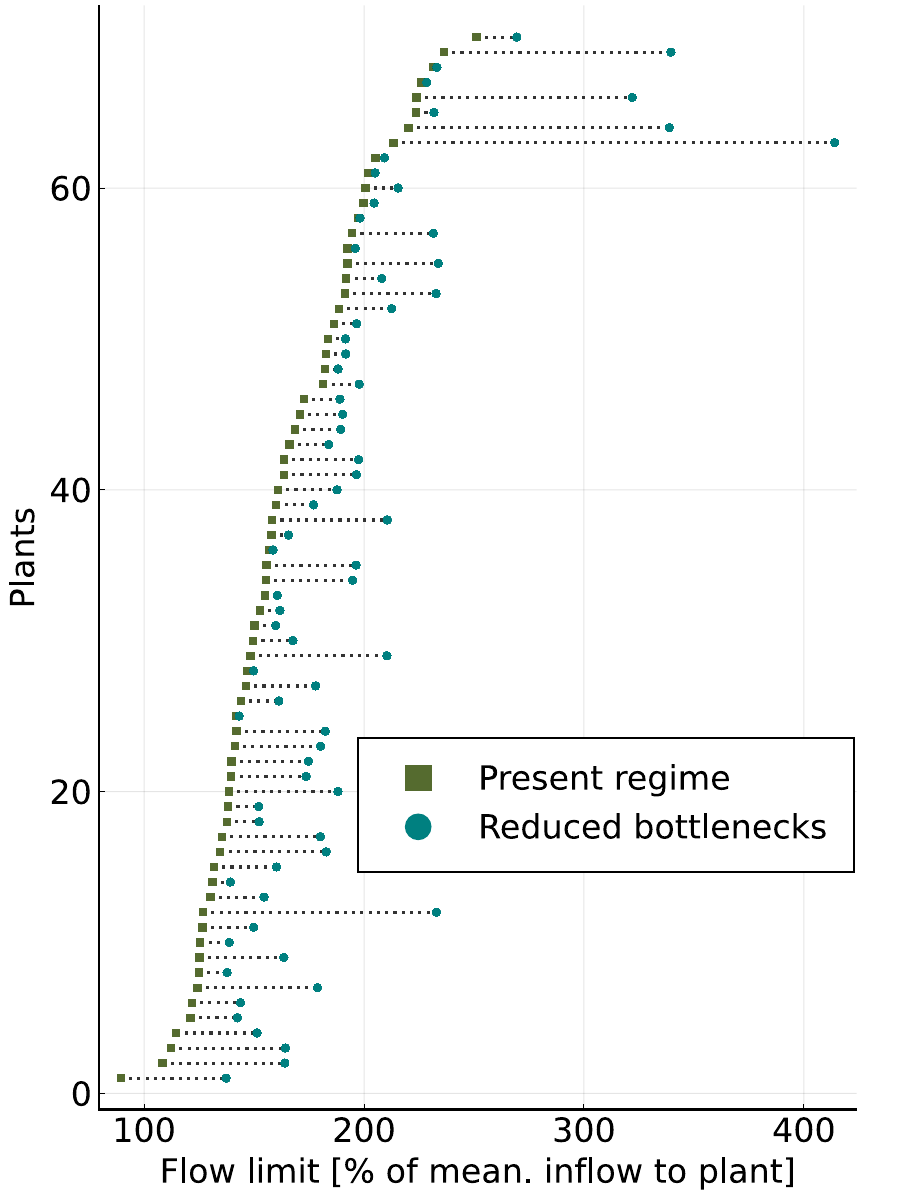}
            \label{fig:expansion_info_mean}
        \end{subfigure}
    \caption{Comparing the flow rate limitations at hydropower bottleneck sites in the \textit{Present regime} and the \textit{Reduced bottlenecks} scenarios. Highlighted are the adjustments made to enhance performance during energy droughts. The left-hand panel displays the flow rate limits as a percentage of maximum inflow to the plant, while the right-hand panel shows these limits as a percentage of the mean inflow to the plant.}
    \label{fig:expansion_info}
\end{figure}

\noindent \textbf{Reduced bottlenecks scenario} \\
In this scenario, we identified all power plants in which the maximum flow capacity of the installed turbines was lower than that of any of the upstream plants. Following this approach, approximately 30\% of all power plants were classified as bottlenecks. For these identified bottlenecks, we increased the allowed short-term regulation [$s$ in Equation (\ref{eq:short})] to match the maximum flow capacity of the turbines in the upstream plant. This adjustment resulted in an average increase of 20\% in the maximum water flows at these plants. When comparing these revised flow rate limits to the maximum recorded inflow for each plant (including the inflow to all upstream plants) over the period of 2016–2020, we found that the limits increased from an average of 18\% to 22\% of the maximum inflow. Similarly, when compared to the average inflow during the same period, the spillage limits rose from an average of 163\% to 194\% of the mean inflow. Details of these adjustments to the flow rate limits, in relation to both the maximum and average inflows, are illustrated in Figure \ref{fig:expansion_info}.

\noindent \textbf{Unrestricted scenario} \\
In the unrestricted scenario, the upper limits on short-term regulation [$s$ in Equation (\ref{eq:short})] were removed for all power plants and reservoirs. Thus, all the plants, including reservoirs without directly connected power plants, were allowed to regulate water without an upper limit. Note that we retained the upper limits on turbine discharge. Thus, we effectively increased the spillage limits.


\subsection{Acknowledgements}
We thank the Swedish water regulation authorities and power production companies for valuable discussions and information regarding Sweden’s river infrastructures and hydropower regulatory frameworks. Special thanks to Anna Hedström Ringvall at Vattenregleringsföretagen and Emma Wikner at Statkraft for their contributions. We also thank Vattenfall, Statkraft, Skelleftekraft, Jämtkraft, Fortum and Energiföretagen for providing the data used in this study.    


\subsection{Declaration of interest}
The authors declare no competing interests.

\newpage
\printbibliography


\appendix
\renewcommand\thefigure{\thesection.\arabic{figure}} 
\setcounter{figure}{0}
\newpage
\section{Supplementary material}

\subsection{Sustained production}

In this section, we present a figure on Sustained production to enhance understanding of sustained output, complementing the Sustained capacity values discussed in the main article.

\begin{figure}[H]
    \centering
        \begin{subfigure}[b]{0.49\textwidth}
            \centering
            \includegraphics[width=\textwidth]{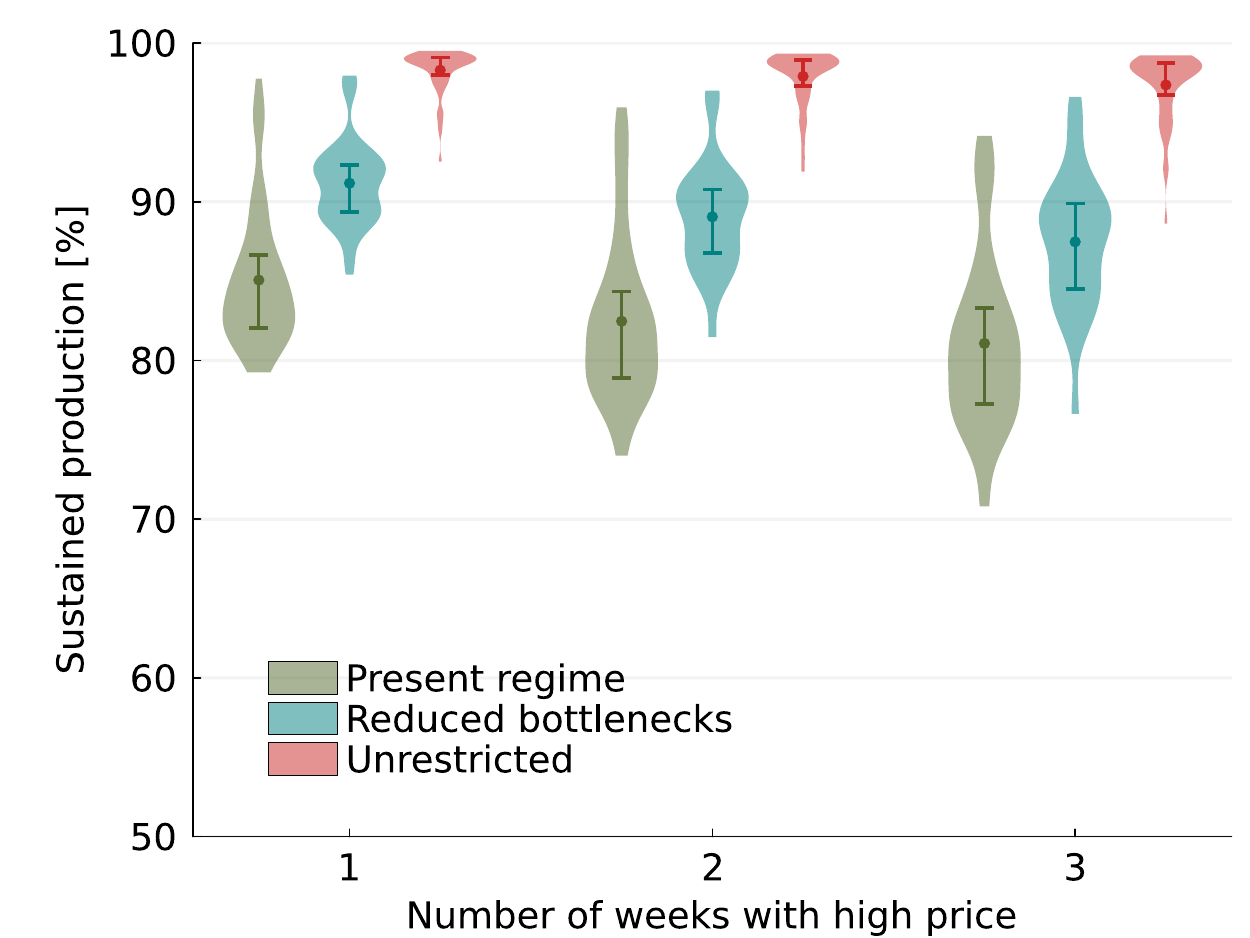}
            \caption{Sustained hydropower production levels in Sweden over a period of one to three consecutive weeks with high electricity prices for three different operational scenarios.}
            \label{fig:A:SO_SP}
        \end{subfigure}
        \hfill
        \begin{subfigure}[b]{0.49\textwidth}
            \centering
            \includegraphics[width=\textwidth]{figures/SE_SC.pdf}
            \caption{This is the same figure as Figure \ref{fig:SC} in the main article. It is reproduced here to facilitate comparison of the Sustained production and Sustained capacity.}
            \label{fig:A:SO_SC}
        \end{subfigure}
    \caption{Illustrations of the sustained hydropower production and the sustained hydropower capacity in Sweden over a period of one to three consecutive weeks with high electricity prices for three different operational scenarios. Each violin contains the results for individual full-year runs with high-price periods for each month over four different years. The dots represent the mean values, while the lines extending from the dots indicate the 25th to 75th percentile range. The color-coding corresponds to the following operational scenarios: \textit{Present regime}, with current regulations and infrastructure (green); \textit{Reduced bottlenecks}, with an increased upper limit on spillage in bottlenecks (blue); and \textit{Unrestricted}, with unrestricted spillage (red). The different scenarios highlight the influence of operational constraints on the ability of hydropower to sustain a high output during energy droughts. The results shown are based on approximately 6500 model runs. Refer to the \textit{Experimental procedure} for details of the test set-up.}
    \label{fig:A:SO}
\end{figure}

\newpage

\subsection{Main results categorised by Nord Pool price areas}

Here, we present the main results from the present study, now categorised according to Nord Pool price areas, and arranged in the same sequence as in the main text.

\subsubsection{Sustained capacity}

\begin{figure}[H]
    \centering
        \begin{subfigure}[b]{0.49\textwidth}
            \centering
            \includegraphics[width=\textwidth]{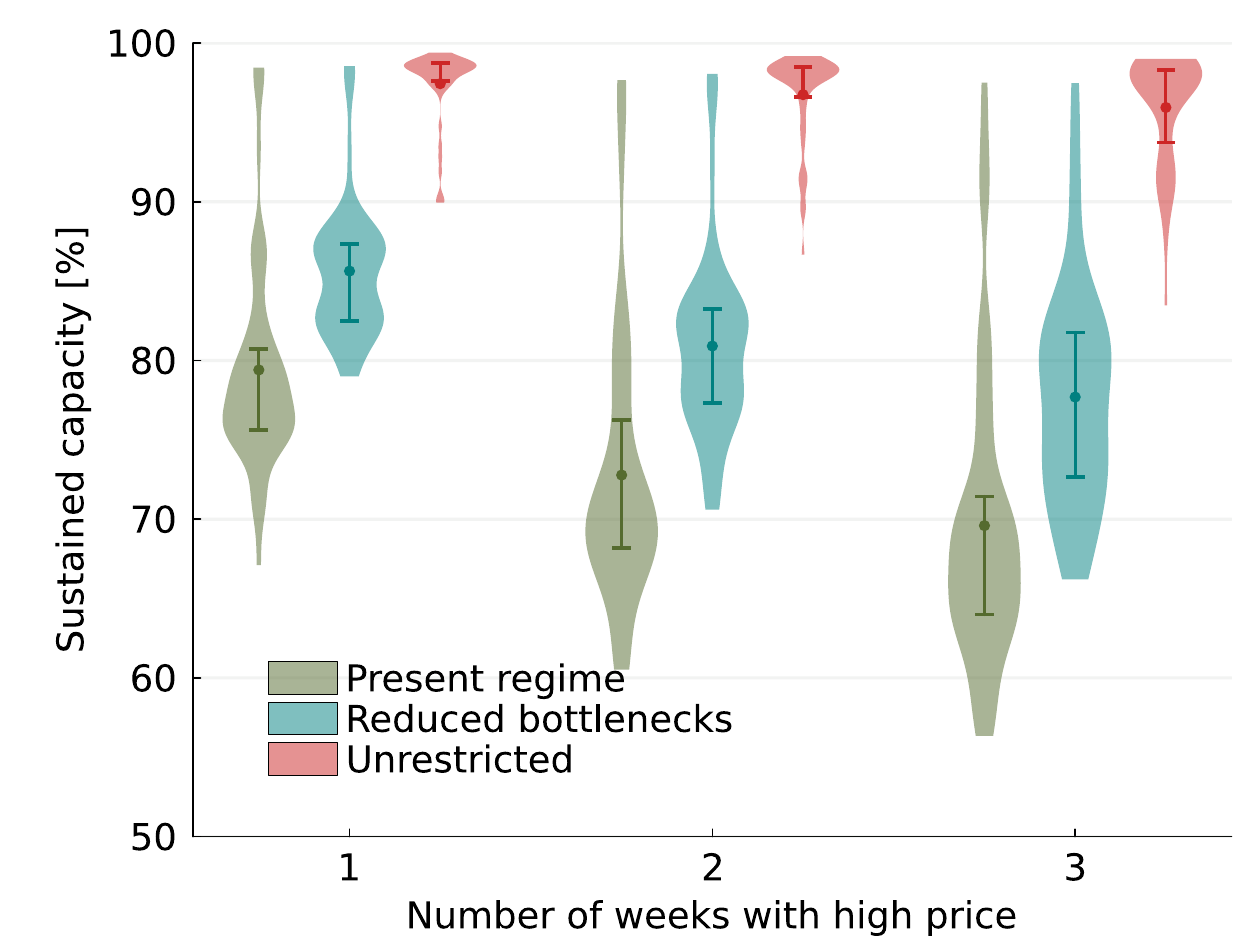}
            \caption{The ability of hydropower in SE1 to sustain a high capacity. The results are based on detailed modelling of 98 \% of the installed capacity in SE1.}
            \label{fig:A:SC_SE1}
        \end{subfigure}
        \hfill
        \begin{subfigure}[b]{0.49\textwidth}
            \centering
            \includegraphics[width=\textwidth]{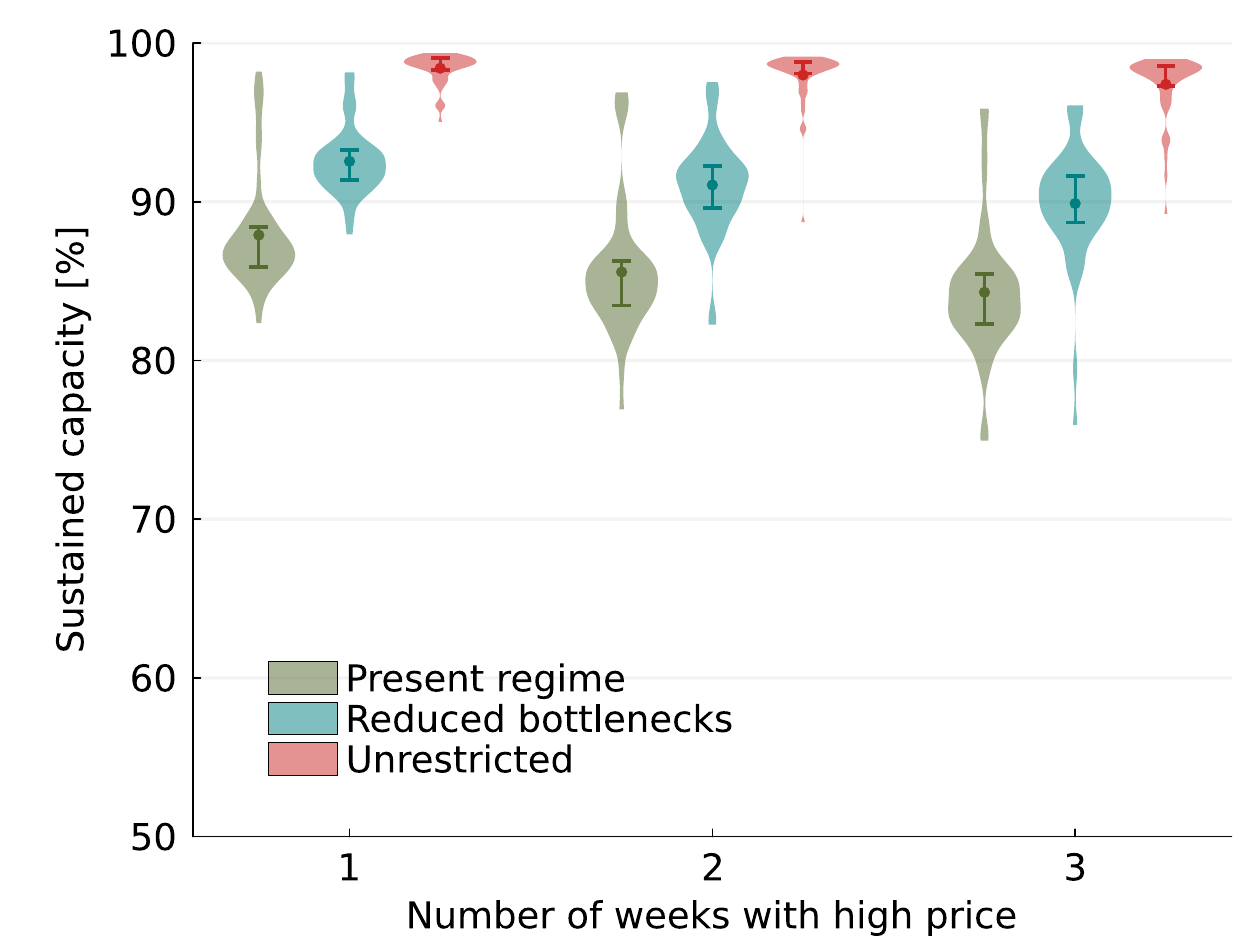}
            \caption{The ability of hydropower in SE2 to sustain a high capacity. The results are based on detailed modelling of 99 \% of the installed capacity in SE2.}
            \label{fig:A:SC_SE2}
        \end{subfigure}
        \vfill
        \begin{subfigure}[b]{0.49\textwidth}
            \centering
            \includegraphics[width=\textwidth]{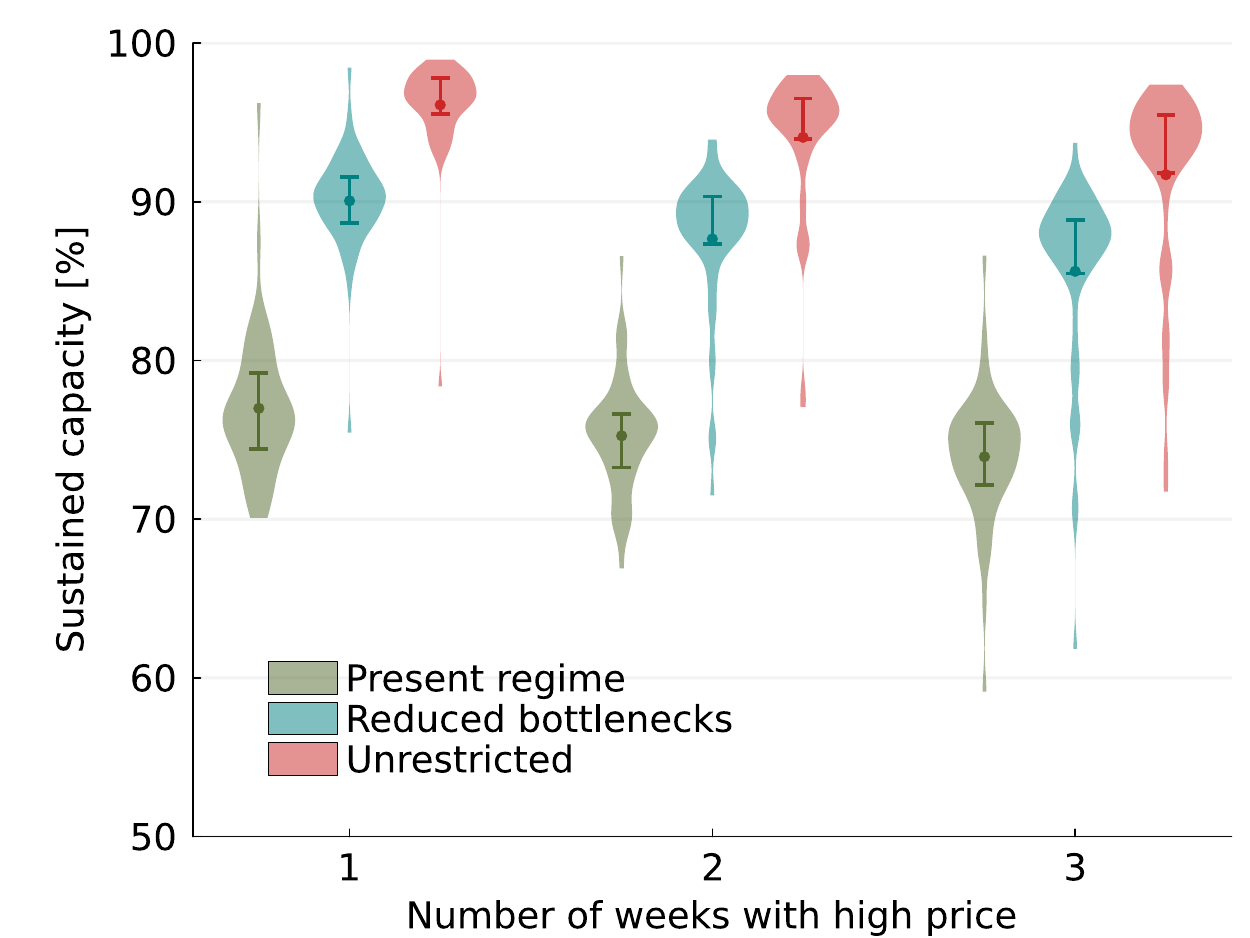}
            \caption{The ability of hydropower in SE3 to sustain a high capacity. The results are based on detailed modelling of 72 \% of the installed capacity in SE3.}
            \label{fig:A:SC_SE3}
        \end{subfigure}
        \hfill
        \begin{subfigure}[b]{0.49\textwidth}
            \centering
            \includegraphics[width=\textwidth]{figures/SE_SC.pdf}
            \caption{The ability of hydropower in SE to sustain a high capacity. The results are based on detailed modelling of 92 \% of the installed capacity in Sweden.}
            \label{fig:A:SC_SE}
        \end{subfigure}
    \caption{Illustrations of the sustained hydropower capacities per price area over a period of one to three consecutive weeks with high electricity prices for three different operational scenarios. Each violin contains the results for individual full-year runs with high-price periods for each month over four different years. The dots represent the mean values, while the lines extending from the dots indicate the 25th to 75th percentile range. The color-coding corresponds to the following operational scenarios: \textit{Present regime}, with current regulations and infrastructure (green); \textit{Reduced bottlenecks}, with an increased upper limit on spillage in bottlenecks (blue); and \textit{Unrestricted}, with unrestricted spillage (red). The different scenarios highlight the influences of operational constraints on the ability of hydropower to sustain a high output during energy droughts. The results shown are based on approximately 6,500 model runs. Refer to the \textit{Experimental procedure} section for details of the test set-up.}
    \label{fig:A:SC}
\end{figure}

\newpage

\subsubsection{Sustained production}

\begin{figure}[H]
    \centering
        \begin{subfigure}[b]{0.49\textwidth}
            \centering
            \includegraphics[width=\textwidth]{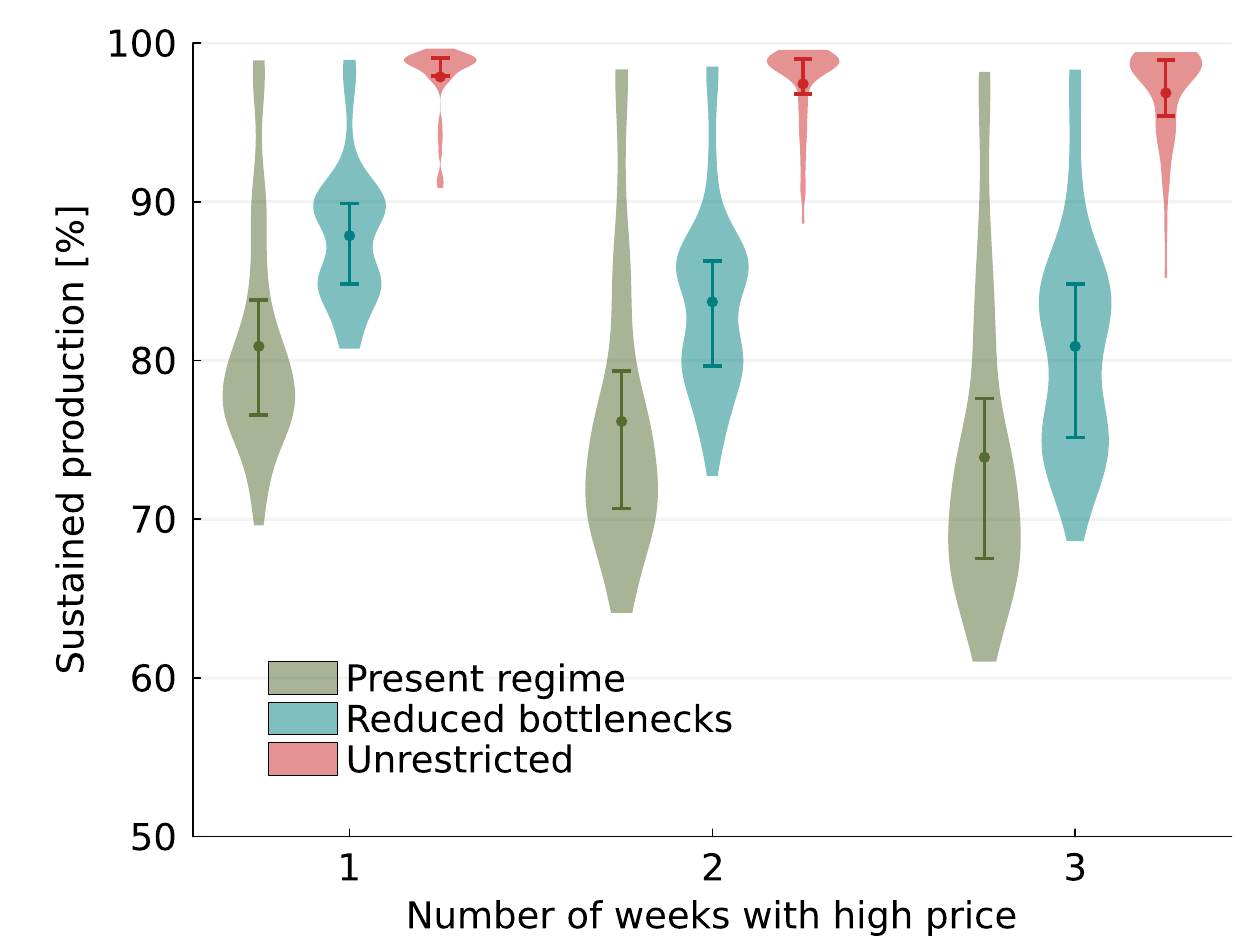}
            \caption{The ability of hydropower in SE1 to sustain high production levels. The results are based on detailed modelling of 98 \% of the installed capacity in SE1.}
            \label{fig:A:SP_SE1}
        \end{subfigure}
        \hfill
        \begin{subfigure}[b]{0.49\textwidth}
            \centering
            \includegraphics[width=\textwidth]{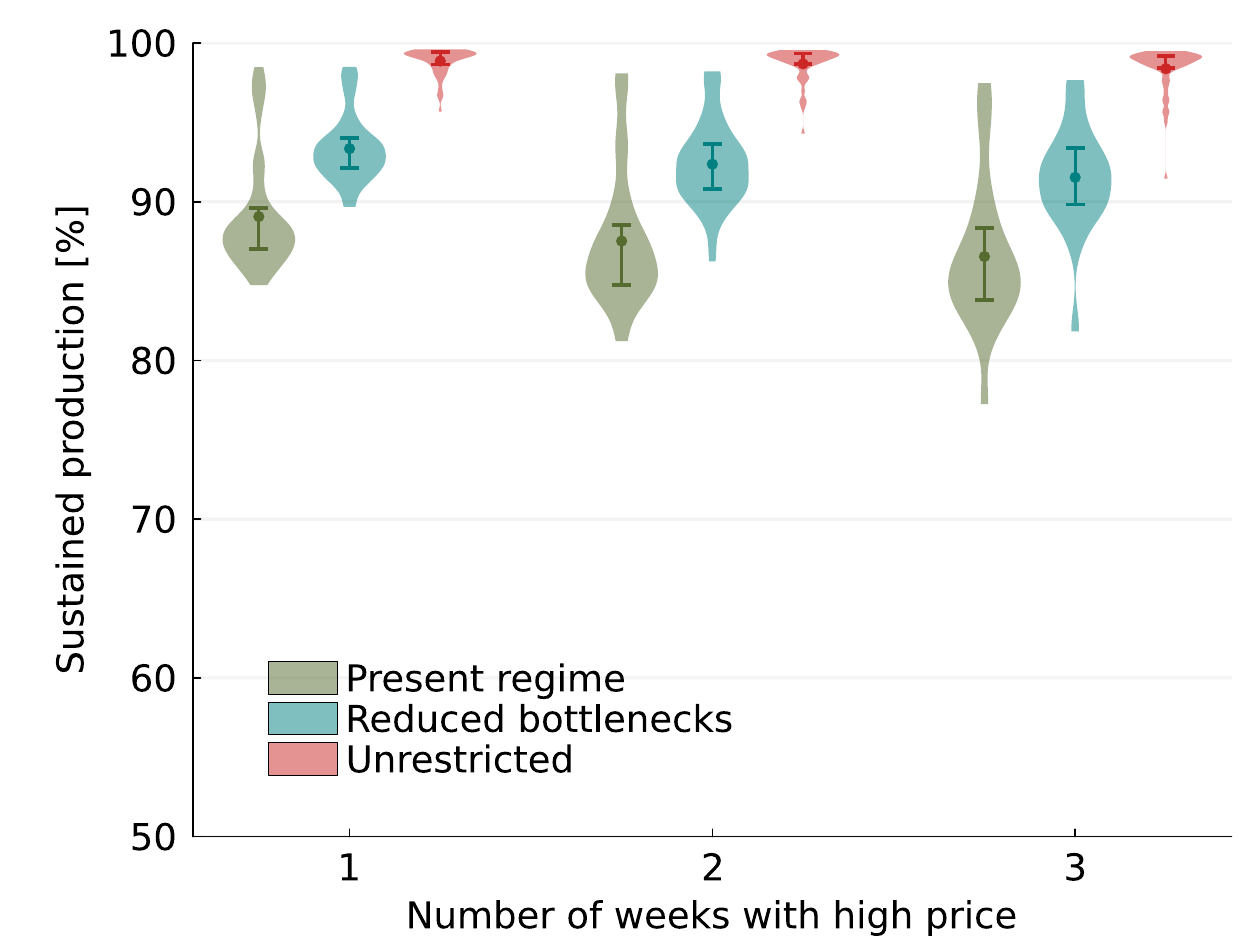}
            \caption{The ability of hydropower in SE2 to sustain high production levels. The results are based on detailed modelling of 99 \% of the installed capacity in SE2.}
            \label{fig:A:SP_SE2}
        \end{subfigure}
        \vfill
        \begin{subfigure}[b]{0.49\textwidth}
            \centering
            \includegraphics[width=\textwidth]{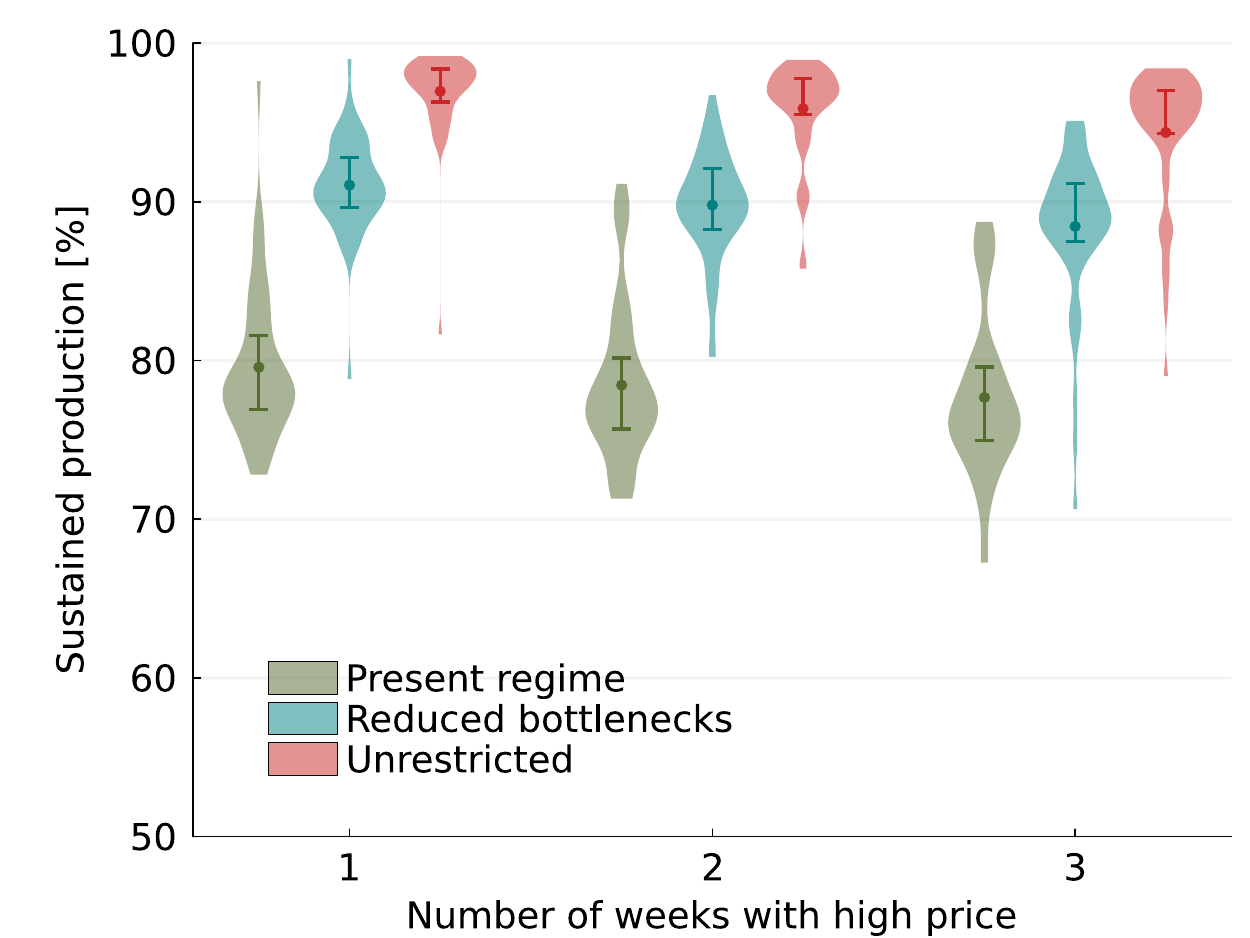}
            \caption{The ability of hydropower in SE3 to sustain high production levels. The results are based on detailed modelling of 72 \% of the installed capacity in SE3.}
            \label{fig:A:SP_SE3}
        \end{subfigure}
        \hfill
        \begin{subfigure}[b]{0.49\textwidth}
            \centering
            \includegraphics[width=\textwidth]{figures/SE_SP.pdf}
            \caption{The ability of hydropower in SE to sustain high production levels. The results are based on detailed modelling of 92 \% of the installed capacity in Sweden.}
            \label{fig:A:SP_SE}
        \end{subfigure}
    \caption{Illustrations of the sustained hydropower production per price area over a period of one to three consecutive weeks with high electricity prices for three different operational scenarios. Each violin contains the results for individual full-year runs with high-price periods for each month over four different years. The dots represent the mean values, while the lines extending from the dots indicate the 25th to 75th percentile range. The color-coding corresponds to the following operational scenarios: \textit{Present regime}, with current regulations and infrastructure (green); \textit{Reduced bottlenecks}, with an increased upper limit on spillage in bottlenecks (blue); and \textit{Unrestricted}, with unrestricted spillage (red). The different scenarios highlight the influence of operational constraints on the ability of hydropower to sustain a high output during energy droughts. The results shown are based on approximately 6500 model runs. Refer to the \textit{Experimental procedure} section for details of the test set-up.}
    \label{fig:A:SP}
\end{figure}

\newpage

\subsubsection{Sustained capacity per month and its correlation with inflow}

\begin{figure}[H]
    \centering
        \begin{subfigure}[b]{0.49\textwidth}
            \centering
            \includegraphics[width=\textwidth]{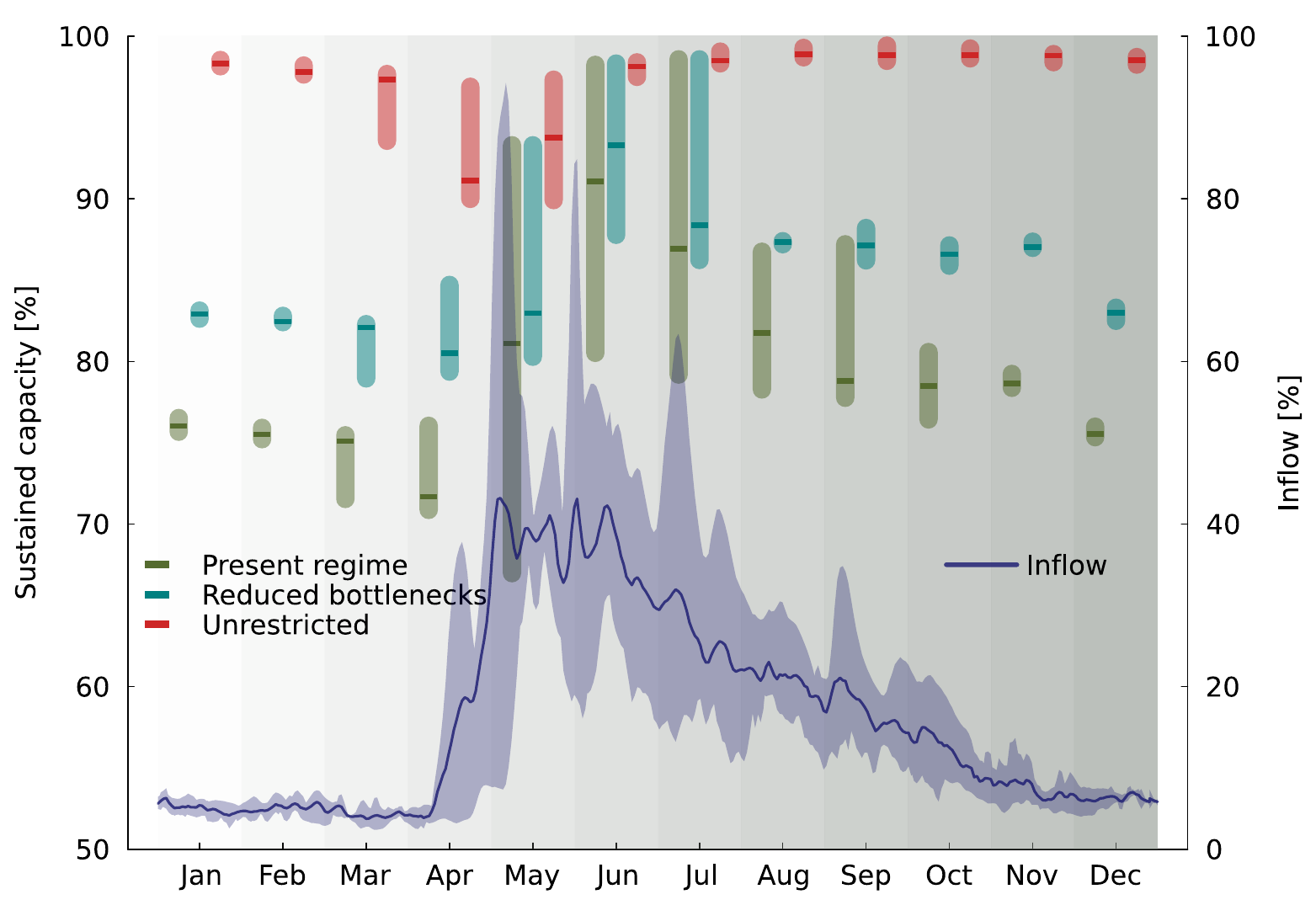}
            \caption{Monthly Sustained capacity in relation to the natural inflow throughout the year in SE1. The results are based on detailed modelling of 98 \% of the installed capacity in SE1.}
            \label{fig:A:SC_month_SE1}
        \end{subfigure}
        \hfill
        \begin{subfigure}[b]{0.49\textwidth}
            \centering
            \includegraphics[width=\textwidth]{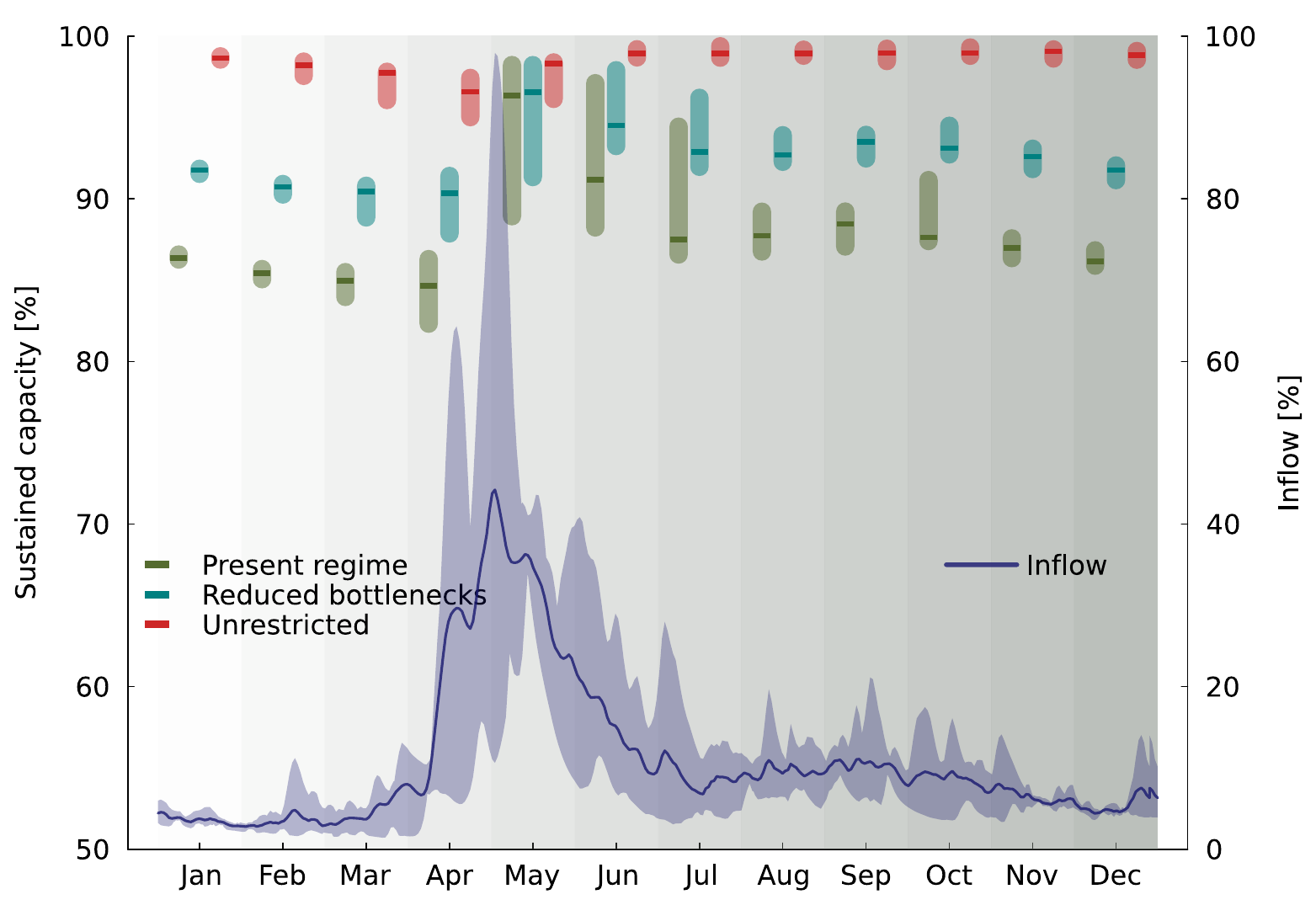}
            \caption{Monthly Sustained capacity in relation to the natural inflow throughout the year in SE2. The results are based on detailed modelling of 99 \% of the installed capacity in SE2.}
            \label{fig:A:SC_month_SE2}
        \end{subfigure}
        \vfill
        \begin{subfigure}[b]{0.49\textwidth}
            \centering
            \includegraphics[width=\textwidth]{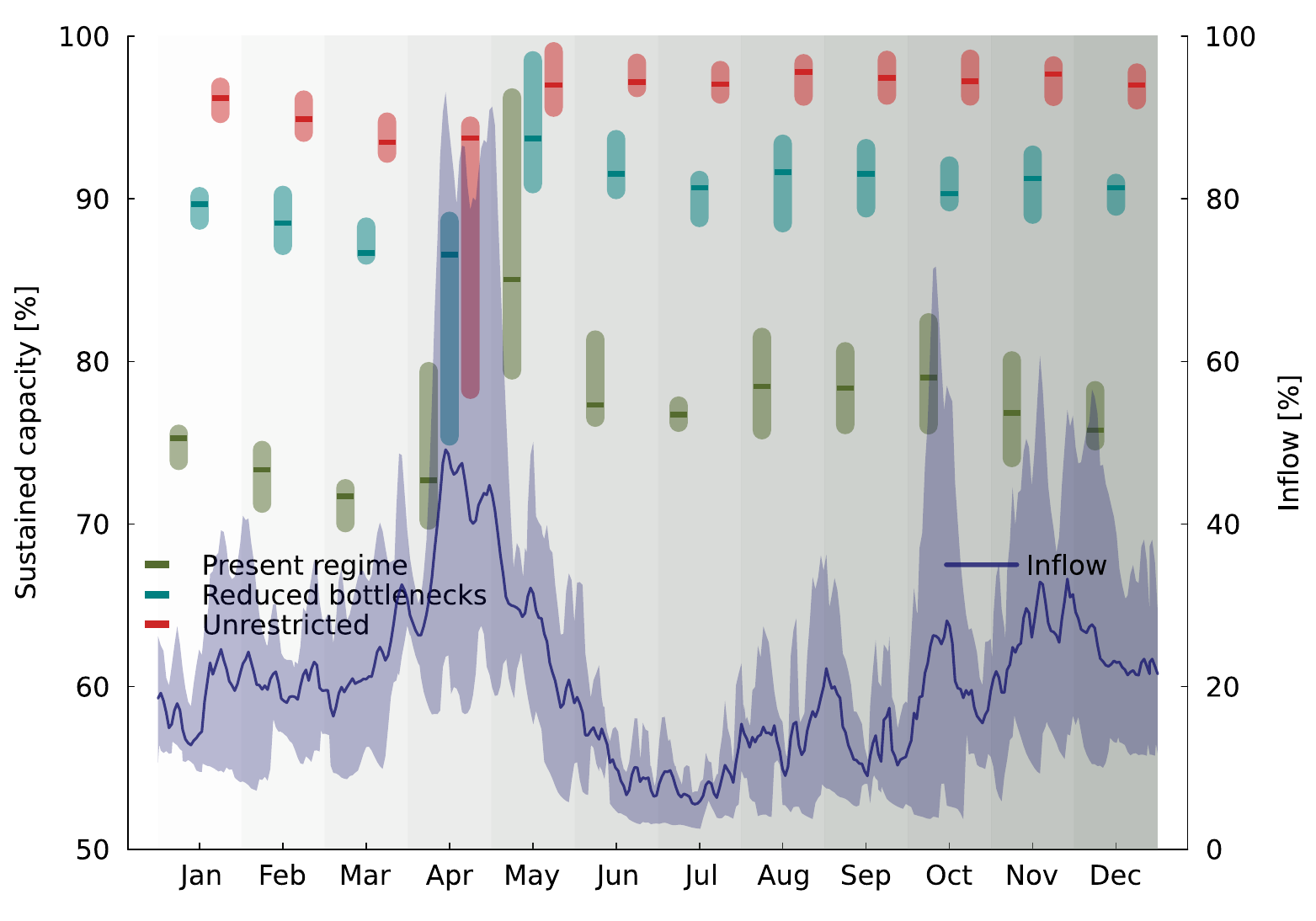}
            \caption{Monthly Sustained capacity in relation to the natural inflow throughout the year in SE3. The results are based on detailed modelling of 72 \% of the installed capacity in SE3.}
            \label{fig:A:SC_month_SE3}
        \end{subfigure}
        \hfill
        \begin{subfigure}[b]{0.49\textwidth}
            \centering
            \includegraphics[width=\textwidth]{figures/SE_SC_month.pdf}
            \caption{Monthly Sustained capacity in relation to the natural inflow throughout the year in all of Sweden. The results are based on detailed modelling of 92 \% of the installed capacity in Sweden.}
            \label{fig:A:SC_month_SE}
        \end{subfigure}
    \caption{A comparison of the monthly sustained capacities of Swedish hydropower in each price area under three operational scenarios versus the natural inflow throughout the year, illustrating how today's hydrological cycle influences hydropower's ability to sustain output. Sustained production levels under the \textit{Present regime} (green), \textit{Reduced bottlenecks} (blue), and \textit{Unrestricted} (red) scenarios are shown through box plots, which display the range and mean values for each month. The line graph represents the total inflow as a percentage of the maximum recorded. The shaded area under the line graph indicates the variability of the inflow in the modelled period of 2016-2020.}
    \label{fig:A:SC_month}
\end{figure}

\newpage

\subsubsection{Loss of total annual production}

\begin{figure}[H]
    \centering
        \begin{subfigure}[b]{0.49\textwidth}
            \centering
            \includegraphics[width=\textwidth]{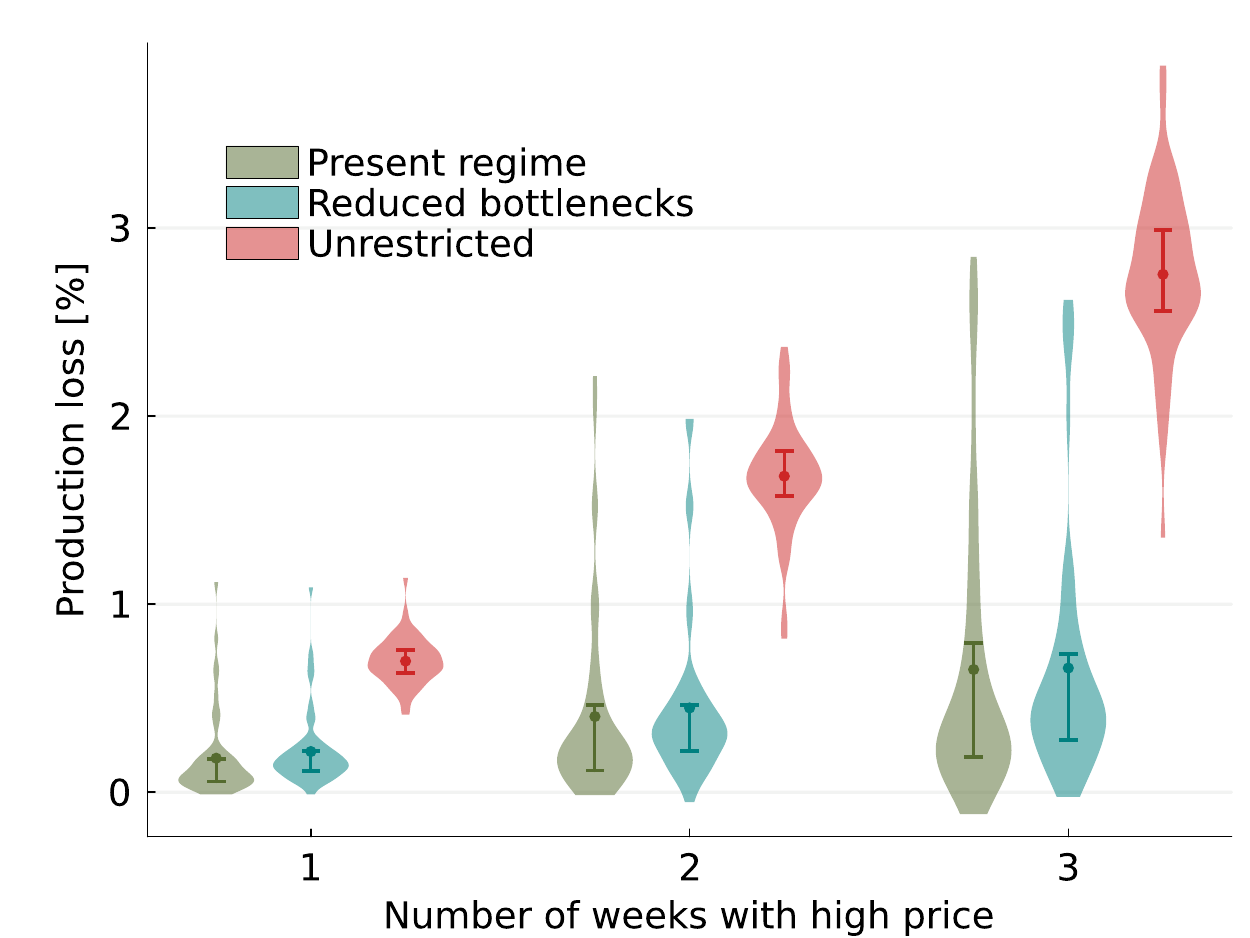}
            \caption{Loss of yearly production induced by sustaining high output in SE1. The results are based on detailed modelling of 98 \% of the installed capacity in SE1.}
            \label{fig:A:Loss_SE1}
        \end{subfigure}
        \hfill
        \begin{subfigure}[b]{0.49\textwidth}
            \centering
            \includegraphics[width=\textwidth]{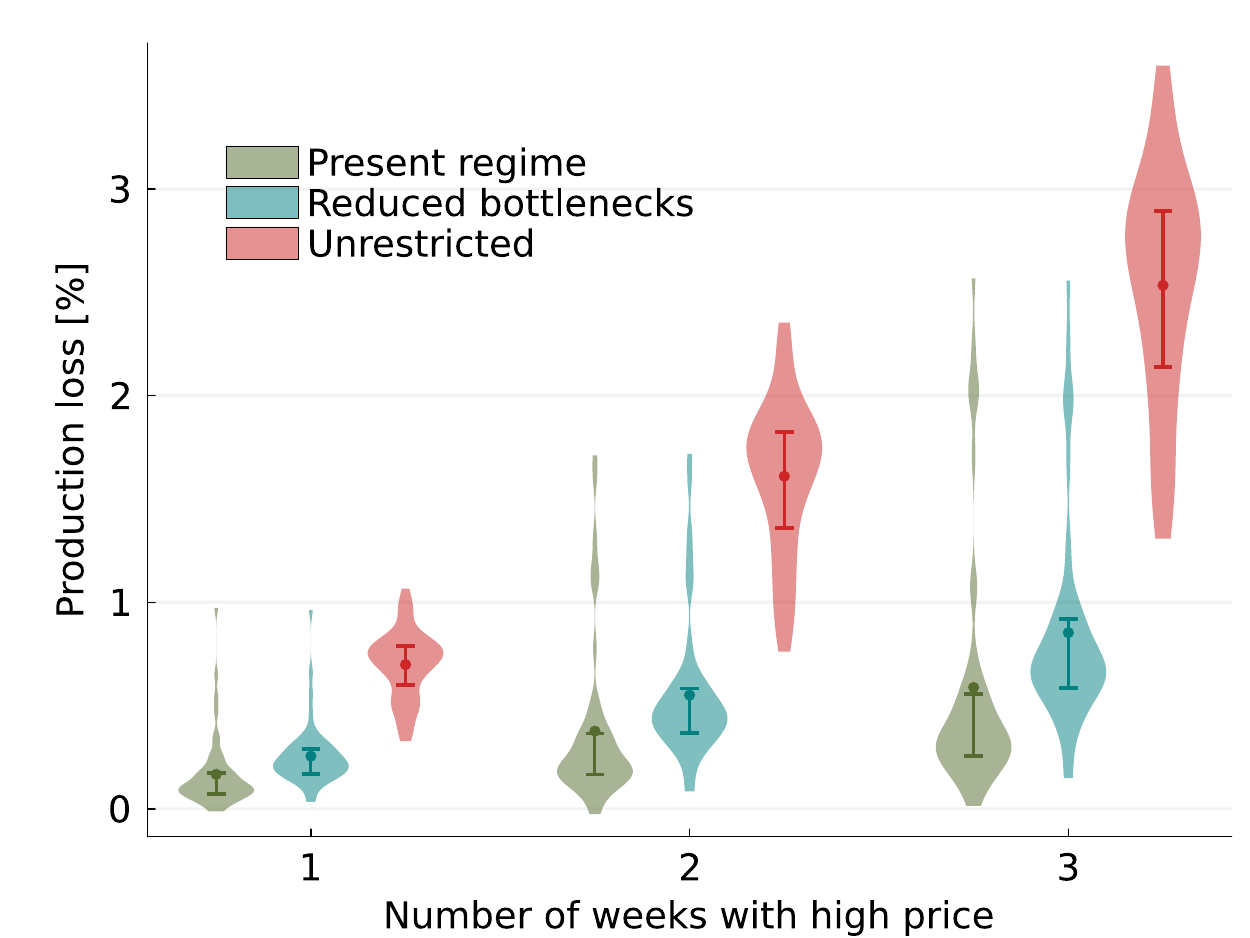}
            \caption{Loss of in yearly production induced by sustaining high output in SE2. The results are based on detailed modelling of 99 \% of the installed capacity in SE2.}
            \label{fig:A:Loss_SE2}
        \end{subfigure}
        \vfill
        \begin{subfigure}[b]{0.49\textwidth}
            \centering
            \includegraphics[width=\textwidth]{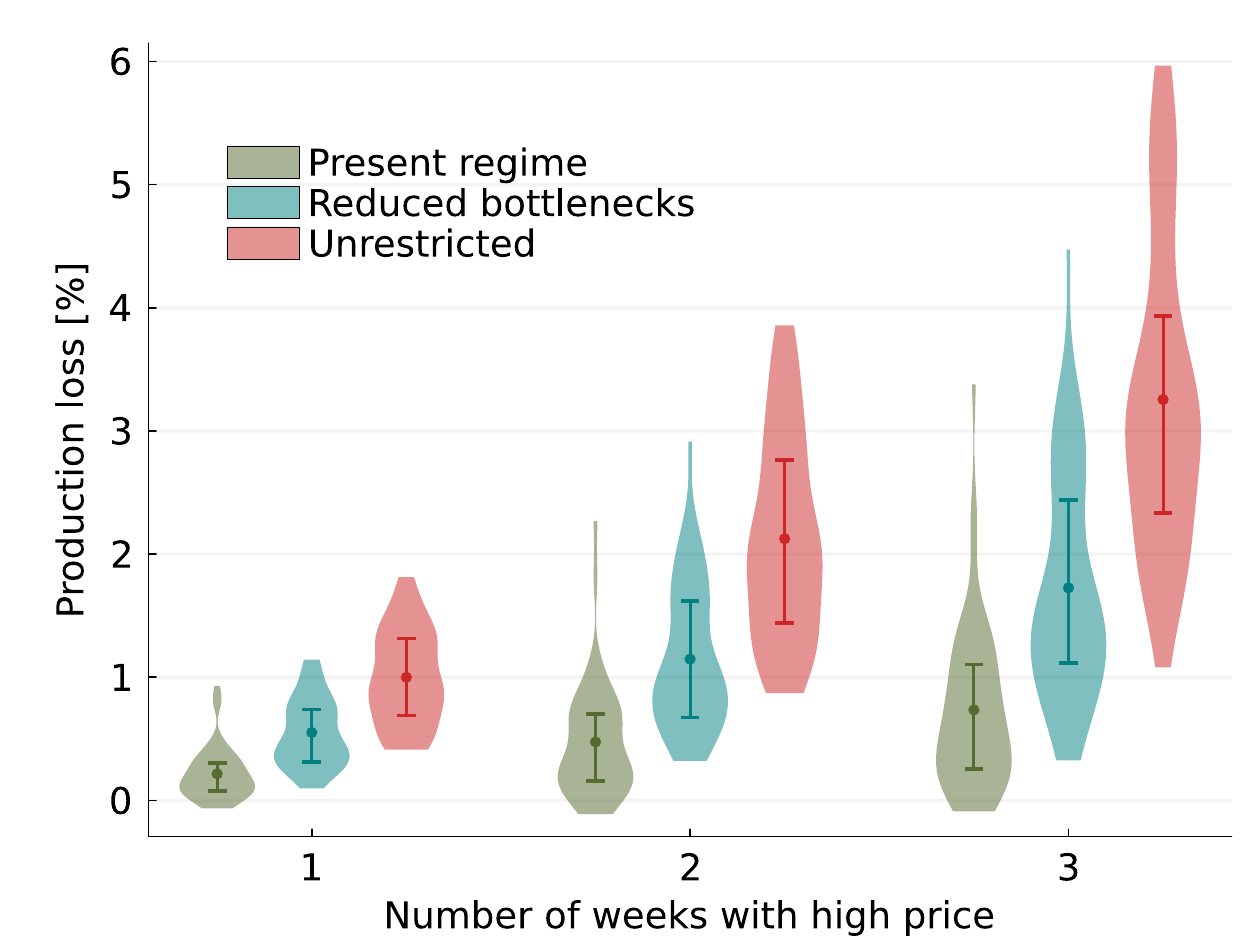}
            \caption{Loss of in yearly production induced by sustaining high output in SE3. The results are based on detailed modelling of 72 \% of the installed capacity in SE3.}
            \label{fig:A:Loss_SE3}
        \end{subfigure}
        \hfill
        \begin{subfigure}[b]{0.49\textwidth}
            \centering
            \includegraphics[width=\textwidth]{figures/SE_Loss.pdf}
            \caption{Loss of in yearly production induced by sustaining high output in all of Sweden. The results are based on detailed modelling of 92 \% of the installed capacity in Sweden.}
            \label{fig:A:Loss_SE}
        \end{subfigure}
    \caption{Visualisation of the losses in yearly production per price area associated with sustaining high output levels, comparing the yearly production levels obtained in the presence of high market prices with the yearly production obtained with historical prices. Each violin contains the results for each month over five different years. The dots represent the mean values, while the lines extending from the dots indicate the 25th to 75th percentile range. The color-coding corresponds to the following operational scenarios: \textit{Present regime}, with current regulations and infrastructure (green); \textit{Reduced bottlenecks}, with an increased upper limit on spillage in bottlenecks (blue); and \textit{Unrestricted}, with unrestricted spillage (red).}
    \label{fig:A:Loss}
\end{figure}

\subsection{Sensitivity analysis}

\begin{figure}[H]
    \centering
    \includegraphics[width=0.5\textwidth]{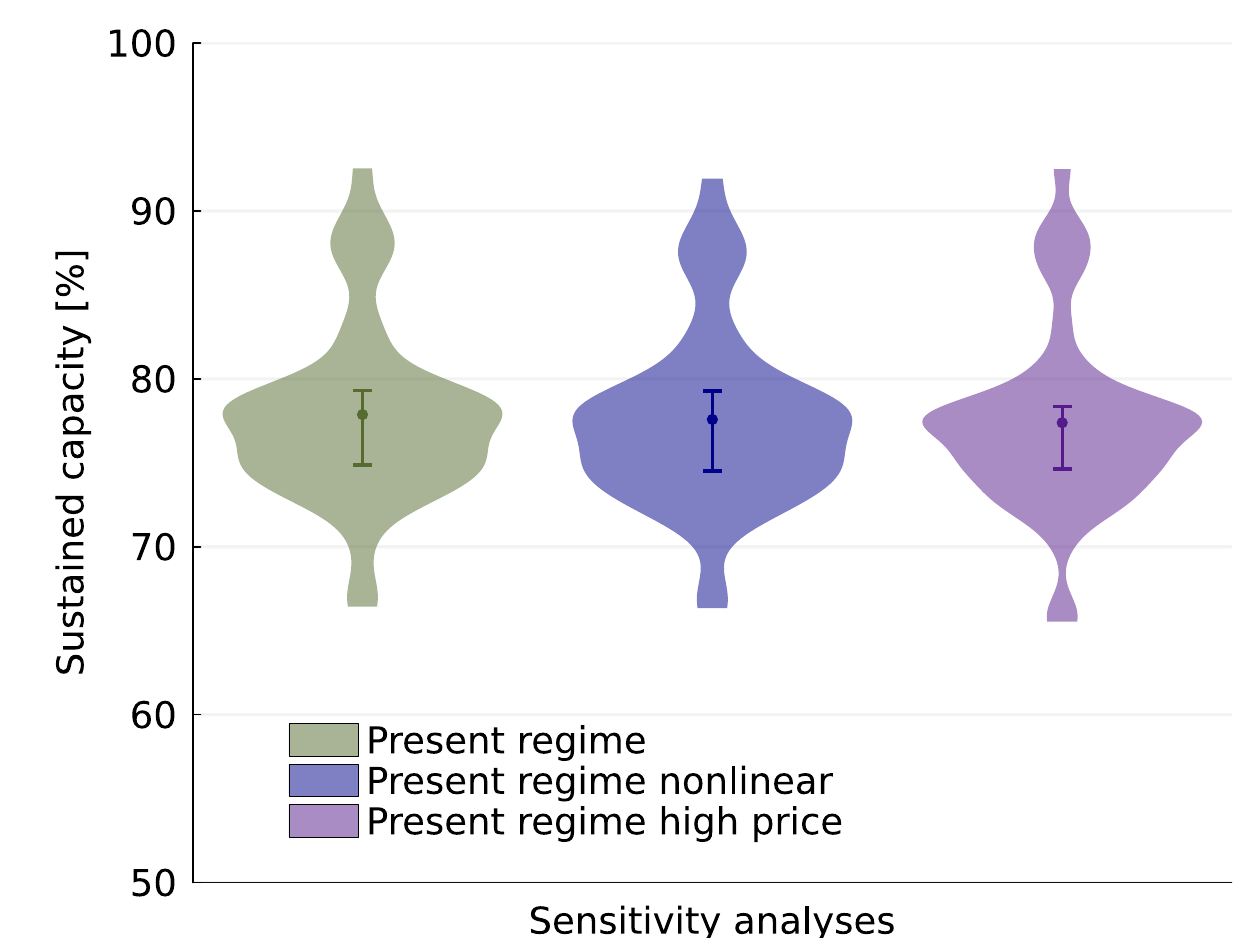}
    \caption{Illustrations of the Sustained capacity in Sweden over three consecutive weeks with high electricity prices. The different violins correspond to the following sensitivity analyses: \textit{Present regime} (left-most violins), with the linearised model (B:L) and a price of 5,000 SEK/MWh (430 €/MWh) during the studied high-price period; \textit{Present regime non-linear} (middle violins), with the full non-linear model (A) and a price of 5,000 SEK/MWh (430 €/MWh) during the studied high-price period; and \textit{Present regime high-price} (right-most violins) with the linearised model (B:L) and a price of 50,000 SEK/MWh (4,300 €/MWh) during the studied high-price period. The \textit{Present regime} violin is the same as that presented as the main result in this study. Each violin contains the results for individual full-year runs with high-price periods for each month over 4 years. The dots represent the mean values, while the lines extending from the dots indicate the 25th to 75th percentile range.}
    \label{fig:A:Sensitivity}
\end{figure}


\end{document}